\documentstyle{l-aa} 

\begin{document}

\newcommand{\equ}{\begin{equation}}
\newcommand{\eequ}{\end{equation}}   
\newcommand{\arry}{\begin{eqnarray}}
\newcommand{\earry}{\end{eqnarray}}
\newcommand{\BF}{\begin{fig[Bure}}
\newcommand{\EF}{\end{figure}}
\newcommand{\BI}{\begin{itemize}}
\newcommand{\EI}{\end{itemize}}
\newcommand{\BE}{\begin{enumerate}}
\newcommand{\EE}{\end{enumerate}}
\newcommand{\dis}{\displaystyle}
\newcommand{\BC}{\begin{center}}
\newcommand{\EC}{\end{center}}
\newcommand{\BL}{\begin{flushleft}}
\newcommand{\EL}{\end{flushleft}}
\newcommand{\BTA}{\begin{table}}
\newcommand{\ETA}{\end{table}}
\newcommand{\BT}{\begin{tabbing}}  
\newcommand{\ET}{\end{tabbing}}
\newcommand{\TAB}{\begin{tabular}} 
\newcommand{\ETAB}{\end{tabular}}
\newcommand{\BD}{\begin{description}}
\newcommand{\ED}{\end{description}}

\newcommand{\ApJ}{{ApJ }}
\newcommand{\AsA}{{A\&A }}
\newcommand{\AsJ}{{AJ }}
\newcommand{\Mn}{{MNRAS }}
\newcommand{\Asp}{{\em Astrophys. Space Sci.}}
\newcommand{\ApJS}{{ApJS }}
\newcommand{\AsAS}{{A\&AS }}
\newcommand{\JQSRT}{{\em J. Quant. Spectros. Radiat. Transfer}}
\newcommand{\ARAS}{{\em Ann. Rev. Astr. Ap.}}
\newcommand{\Via}{{\em Vistas in Astronomy}}
\newcommand{\ApJL}{{\em Astrophys. J. Lett.}}
\newcommand{\Na}{{\em Nature}}
\newcommand{\AnA}{{\em Ann. d'Ap.}}
\newcommand{\PhR}{{\em Phys. Rev.}}
\newcommand{\Nse}{{\em Nucl. Sci. Engng.}}
\newcommand{\Msait}{{Mem. Soc. Astron. Ital. }}

\newcommand{\Spe}{{Spectral Evolution of Galaxies}}
\newcommand{\RT}{{\em Radiative Transfer}}
\newcommand{\Ntt}{{\em Neutron Transport Theory}}
\newcommand{\Ppim}{{\em Physical Processes in the Intestellar Medium}}
\newcommand{\Iaunu}{{\em IAU Symp.\ 1991}}   
\newcommand{\Iauoq}{{\em IAU Symp.\ No.\ 108}}
\newcommand{\Rmcr}{{\em Recent developments of Magellanic Cloud Research}}
\newcommand{\Sn}{{\em Stellar Nucleosynthesis}}
\newcommand{\Iauon}{{\em IAU Symp.\ No.\ 135 1989, }}
\newcommand{\NIM}{{\em Nebulae and Interstellar Matter}}
\newcommand{\Iauno}{{\em IAU Symp.\ No.\  1990, The Galactic and Extragalactic}}
\newcommand{\Lssp}{{\em Light Scattering by Small Particles}}
\newcommand{\AGAS}{{\em Astrophysics of Gaseous Nebulae and Active Galactic
Nuclei}}
\newcommand{\PGAS}{{\em Physics of Thermal Gaseous Nebulae}}
\newcommand{\ND}{{\em Numerical Data and Functional Relationships in Science
and Technology}}
\newcommand{\nrsc}{{\em New Results on Standard Candles}}
\newcommand{\RMAA}{{\em Rev. Mex. A.A.}}
\newcommand{\GEO}{{\em Geochim. et Cosmochim. Acta}} 
\newcommand{\Pasp}{{PASP }}

\thesaurus{3(11.05.2; 11.19.5; 11.06.2; 09.08.1)}

\title{The determination of the star formation rate in galaxies}

\author{G.~Barbaro \inst{1}
\and B.M.~Poggianti
\inst{2,3,4} }

\offprints{G.~Barbaro}

\institute{Dipartimento di Astronomia, vicolo dell'Osservatorio 5, 35122
Padova, Italy, barbaro@astrpd.pd.astro.it \and
Institute of Astronomy, Madingley Road, Cambridge CB3 0HA, UK \and
Royal Greenwich Observatory, Madingley Road, Cambridge CB3 0EZ, UK
\and Kapteyn Instituut, P.O. Box 800, 9700 AV Groningen, The Netherlands,
bianca@astro.rug.nl}

\date{Received; accepted}

\maketitle

\begin{abstract}

A spectrophotometric model able to compute the integrated spectrum of a galaxy,
including the contribution both of the stellar populations and of the ionized
interstellar gas of the HII regions powered by young hot stars, has been used
to study several spectral features and photometric quantities in order to
derive calibrations of the star formation history of late type galaxies.
Attention has been paid to analyze the emission of the Balmer lines and the
[OII]$\lambda$3727 line to test their attitude at providing estimates of the present
star formation rate in galaxies. Other features, like D$_{4000}$ and the 
equivalent width of the H$_{\delta}$ line, influenced by the presence
of intermediate age stars, have been considered.

Several ways of estimating the star formation rates in normal galaxies are
discussed and some considerations concerning the applicability of the models
are presented. Criteria have been also studied for ascertaining the presence of
a burst, current or ended not long ago. Bursts usually hinder the determination
of the past star formation rate. 


\end{abstract}

\section{Introduction}

The star formation history is a fundamental quantity in the study of the galaxy
populations and their evolution. It is becoming evident that this evolution
can be influenced by the interactions with the intergalactic medium or with
other galaxies, giving rise to episodes of intense star formation (bursts).
Accordingly the elaboration of methods for evaluating the star formation rate 
(SFR) is one of the
tasks of the research in this field. 
 
Clearly the needed information arises from the integrated spectra and
especially from spectral features contributed by stars or by the ionized
interstellar gas. In this work we present an analysis of some of such features
which are able to disclose the presence of stars of a given age and
therefore can yield estimates of the SFR in different epochs of the galaxy
life. To this aim we have developed a spectrophotometric model which takes
into account
both the stellar and the nebular contribution. The model predictions have been
compared with a sample of observational data and from this comparison suitable
quantities for the evaluation of the SFR have been singled out. 
Moreover criteria for
ascertaining the occurrence of bursts, both current or recently concluded, have
been derived. 

The observational material consists of the sample of spectra of local galaxies
of Kennicutt (1992a) including both normal galaxies, that is galaxies of the
standard morphological sequence, and starburst galaxies. With such spectra,
available in digital records from the Astronomical Data Center, Kennicutt's
analysis has been extended by deriving other spectral features besides those
computed by him. The comparison with the spectrophotometric model has
allowed: a) to derive calibrations of the SFR in terms of several quantities,
especially the intensities of the emission lines, b) to single out 
relations which are valid independently of the history of star formation of
the galaxy, c) to find out quantities which can descriminate between normal 
and starburst galaxies. 

The comparison with the spectra of normal galaxies allows to test the
capability of the model to reproduce a large set of spectral features for all
the galactic types. The model thus tested can be used generally in the 
study of distant galaxies in the frame of the problems of galaxy evolution. 
 
\section{The spectrophotometric model}

The theoretical tool used is an evolutionary synthesis model able
to compute the integrated spectrum of a galaxy from the far UV region up to the
IR including both the contribution of the stellar component and the thermal
emission of the gas in HII regions. The non-thermal gaseous emission, 
arising from
non-stellar ionizing sources and important only in AGN, and the far IR emission
of dust are not considered. 

\subsection{The stellar spectrum}

The emission of the stellar component is derived with an updated version of a
previous model computed by Barbaro and Olivi (1986, 1989). It takes into
account all the advanced evolutionary phases up to the AGB and post-AGB, while
it does not consider the pre-main sequence, where the visible light of the most
luminous stars is completely extincted or their contribution is negligible due
to the small lifetime of this phase. The presence of populations of stars of
different metallicity is envisaged, the chemical evolution being computed in a
simple way. While the evolutionary background, which has been described in 
detail in Barbaro \& Olivi (1989), has not been changed, substantial
improvements have been introduced in the spectral library. First of all the new
atmosphere models of Kurucz (in the 1993 version) have been adopted. 

In the IR region atmosphere models of cool stars are not much reliable because
of the uncertainties affecting the input of atomic and molecular data
concerning the opacities. Therefore for stars with $T_{eff}<5500$ K the
library of observed spectra of Lan\c{c}on - Rocca Volmerange (1992, LRV) has
been adopted. This library covers the spectral interval 14500 - 25000 \AA $\,$
(and therefore includes the H and K bands of Johnson's photometric system) with
a resolution of about 25 and 70 \AA $\,$ respectively at the lower and the
upper end of the observed region. Although it includes stars with different
metallicities, the small number of them prevents an organization in categories
of different metal content and therefore a unique calibration independently of
Z has been adopted. For stars with $T_{eff}> 5500$ K Kurucz's models
have been used. 
The region of wavelengths higher than 25000 \AA $\,$ has not been
included as it is heavily affected by the dust emission, which 
we have not considered.
A crucial point is the fit of the visible and IR spectra: the fit has 
been performed with the aid of a black body
spectrum. The difficult task is to determine its temperature due to the weak
dependence on it of the spectrum at such wavelengths. We have adopted an
iterative procedure by choosing a suitable temperature and using the
corresponding black body as a fitting support; with the obtained spectrum broad
band optical and IR colours have been obtained and compared with those observed
for the same effective temperature (Koornneef 1983, Bessell \& Brett 1988,
LRV). The fitting temperature was changed until agreement between computed and
observed colours was reached. Kurucz's models and observed spectra of LRV agree
very well for early type and intermediate type stars and on this ground
theoretical spectra have been used also for sufficiently hot stars. Systematic
differences appear instead for later types, for which the computed (H-K) colour
is bluer than the observed one. 

Kurucz's models (1993) have been also used for the synthesis of the integrated
spectra in the far UV region. From the spectral energy distributions in the UV
region the flux of the ionizing photons has been derived in the computation of
the nebular emission. The spectral distributions of stars hotter than 50000 K,
not included in Kurucz's library, have been approximated with black body
spectra. 

Since the spectral resolution of Kurucz's atmosphere models does not 
allow the synthesis of some spectral features  (e.g. the Balmer absorption
lines) in the visible region, also the library of stellar spectra of Jacoby et
al. (1984) has been used. The stars considered in this last catalogue cover
all the spectral types from O to M, all the luminosity classes and have, with
few exceptions, solar metallicity; their spectra, which have been corrected
for interstellar reddening, have a resolution of about 4.5 \AA. The link with
the isochronous lines has been obtained with Schmidt-Kaler's $T_{e}$-- 
Spectral Type
relations, as presented in tables 4.1.4.3 and 4.1.5.23 of Landolt -
B\"ornstein (1982). In the common spectral interval both libraries give almost
equal integrated continua of simple stellar populations and galaxies. Small
differences appear when comparing spectra of metal--poor populations since, as
already remarked, the spectra of Jacoby et al. (1984) have solar composition;
however, as long as extreme situations are excluded, the results are very
similar. 

\subsection{The ionized gas emission: the line spectrum}

Models of HII regions excited by single stars of different effective
temperature have been computed by several authors: in particular Rubin (1985)
and Stasinska (1990) have investigated a variety of parameters (chemical
composition and density of the gas, effective temperature and luminosity of the
exciting stars). From the analysis of these models a relevant feature emerges:
in models with different ionization parameter $U$ the line fluxes do not scale
with the flux of the ionizing photons. This implies that the emission of an HII
region powered by a group of stars, all having the same temperature, cannot be
computed from the models of an HII region powered by a single star having that
temperature and luminosity equal to the sum of luminosities. Having in mind
these circumstances let us examine the construction of the nebular spectrum of
a galaxy. Let us assume that the nebular emission of the galaxy is the sum of
the contributions of the HII regions excited by all the hot young stars. The
integrated $H{\beta}$ luminosity, in the hypothesis that the HII regions are
ionization-bound, is then given by: 

\equ
L(H \beta)= {\dis \int \int} B(m,t) l_{H \beta}(m,t) dm dt
\label{for101}
\eequ

where $l_{H \beta}(m,t)$ is the  $H{\beta}$ luminosity of a star of mass $m$
 born
at time $t$. $B(m,t)$ is the distribution function of masses and birth times of the
stars. In principle the integral should be extended to all the hot stars with a
conspicuous UV flux, but in the case of late type galaxies, such as
those of concern in the present work, the ionizing radiation due to the young
massive stars is at least five order of magnitude larger than that originating
from hot old stars (Binette et al. 1994). According to Osterbrock (1989), 
$l_{H \beta}(m,t)$ is given by: 

\equ
l_{H \beta}(m,t)= h \nu_{H \beta} \frac{\alpha_{H \beta}}{\alpha_{B}} 
n_{c}(m,t)=K_{H \beta} n_{c}(m,t)
\label{for102}
\eequ
where  $ h \nu_{H \beta}$ is the energy of photons with $\lambda=4862$ \AA,
$\alpha_{H \beta}$ and  $\alpha_{B}$ are respectively the coefficient of
recombination  with emission of the $H\beta$ line and the total recombination
coefficient (case B, optically thick). $K_{H \beta}$ is a function of the
electron temperature which, to a first approximation, depends only on the
chemical  composition. $n_{c}$ is the number of ionizing photons emitted per
second from a star with mass $m$ and age $\tau=T_{G}-t$, where $T_{G}$ is the age
of the galaxy. By keeping $K_{H \beta}$=constant, equation (\ref{for101})
becomes: 

\equ
L(H_{\beta})= K_{H \beta} {\dis \int \int} B(m,t) n_{c}(m,t) dm dt=
 K_{H \beta} N_{c}
\label{for103}
\eequ
being $N_{c}$ the integrated ionizing flux of the galaxy,

\equ
N_{c}= {\dis \int \int} B(m,t) n_{c}(m,t) dm dt
\eequ

Usually one sets: $ B(m,t)= \psi(t) \phi(m)$ and therefore:

\equ
N_c= {\dis \int_{T-{\tau}_M}^T} \psi(t) {\cal N}_c(T-t) dt
\eequ

which is computed by considering only hot stars with $\tau \le \tau_{m}=  4
\cdot 10^7$ yr; ${\tau}_{M}$ is the maximum age of a cluster still able to
power a non negligible HII region. ${\cal N}_c(T-t)$ is the integrated UV flux
of a generation born at the time $t$.
 
This procedure can be extended to a generic Balmer line $H_{\iota}$ and to any
other hydrogen emission line since: 

\equ
\frac{l_{H \iota}(m,t)}{l_{H \beta}(m,t)}= h_{H \iota}
\label{for105}
\eequ
(Osterbrock 1989)  with $h_{H \iota}$ is function of the gas temperature alone.
The independence of $l_{H \iota}/l_{H \beta}$ from $m$ and $t$ can be verified
on Rubin's (1985) and Stasinska's models (1990). Therefore:

\equ
L(H_{\iota})= h_{H \iota} {\dis \int \int} B(m,t) l_{H \beta}(m,t) dm dt=
 h_{H \iota} K_{H \beta} N_{c}
\label{for106}
\eequ   

The behaviour of the emission lines of elements other than hydrogen
is different. Let us consider the line of the generic $X$  element
(such is for instance $OII$) at the wavelength $\lambda$. We then have:

\equ
L(X_{\lambda})= {\dis \int \int} B(m,t) l_{x_{\lambda}}(m,t) dm dt
\label{for107}
\eequ   
with

\equ
\frac{l_{x_{\lambda}}(m,t)}{ l_{H\beta}(m,t)}=h_{X_{\lambda}}(m,t)
\label{for108}
\eequ

According to the models of Rubin (1985) and Stasinska (1990) 
$l_{x_{\lambda}}/ l_{H\beta}$ is really function not only of the effective 
temperature of the source of the ionizing photons but also of their luminosity
and therefore of $m$ and $t$, in the frame of the theories of stellar evolution.
Therefore the line luminosity is given by
 
\equ
L(X_{\lambda})= {\dis \int_{T-{\tau}_M}^T} \psi(t) {\cal L}_{X_{\lambda}}(T-t) dt
\label{for109}
\eequ
where ${\cal L}_{X_{\lambda}}(T-t)$ is the contribution of all the stars born at
the time $t$. No other simplification can be introduced. 

We can therefore conclude that: 

a) as far as the Balmer lines are concerned, by virtue of equation
(\ref{for103}), a galaxy can be treated as a unique HII region powered by a
source having a flux of ionizing photons equal to the flux of all the young hot
stars. The same conclusion applies also to the other H series. Therefore 
equations (\ref{for103}) and (\ref{for106}) have been adopted.
 
b) on the contrary the same simplification is not possible for the metallic
lines and for such lines the integrated emission of the galaxy must be computed
as the sum of the contributions of all the HII regions present, as consequence
of eq. (\ref{for109}). To this purpose we can imagine the galaxy as a complex
of several HII regions each powered by a cluster of stars. This model is
realistic since we observe that the majority of massive stars are born in groups.
Depending on the SFR, and therefore on the number of young clusters, two
situations can occur:  a) the HII regions do not overlap each other: in this
case the nebular emission of the galaxy is the sum of the contributions of all
the HII regions; b) the HII regions partly or totally overlap: here the
situation is more complex, but the extreme case can be adopted in which all
the hot young stars of the galaxy power a unique large HII region. With a SFR
equal to that actually measured in the disc of our galaxy and assuming a
cluster mass of 3000 $M_{\odot}$ the sum of the volumes of all the HII regions
is by several orders of magnitude smaller than the disc volume: the assumption
that there is no overlap seems therefore a correct approximation for normal
galaxies. Very likely only for mergers with a larger SFR a more correct model
would be that of a single large HII region powered by all the young stars. 
The relations giving the luminosity of the H lines are still valid even in the
case of overlap of the HII regions, since they imply only the total UV flux.

For each cluster of age $\tau$ the integrated luminosity 
${\cal L}_{X_{\lambda}}(\tau)$ of the ${X_{\lambda}}$ line has been obtained by approximating
the integrated stellar spectrum with the spectrum of a single star with
temperature $T_{eq}$ and luminosity equal to the luminosity of the cluster. The
equivalent star temperature has been determined by fitting the portion  of the
integrated spectrum of the cluster in the range  300--1000 \AA $\,$ to the
corresponding portion of the stellar spectrum. The lower limit of such interval
has been arbitrarily fixed at  300 \AA $\,$. For smaller wavelengths the
integrated spectrum is dominated by stars hotter than 50000 K, which is the
upper limit of the temperatures of Kurucz's models. These stars are treated as
black bodies, but this approximation is unsatisfactory and when the fitting is
extended to $\lambda \le$  300 \AA $\,$ too large equivalent temperatures
ensue. It has been shown in fact with the hottest Kurucz' models that the
monochromatic flux derived by the Plank law, for $\lambda < 300$  \AA $\,$, is
larger than  that of the corresponding model. 

The number of equivalent stars (i.e. those giving on the whole the same 
luminosity of the cluster) is equal to the ratio of photons emitted by the 
cluster stars to the photons emitted by a main sequence (MS) star 
with $T_{eff}=T_{eq}$.
The calibration of Panagia (1973) has been used for this purpose. One then 
gets:

\equ
{\cal L}_{X_{\lambda}}(\tau)= h_{X_{\lambda}} {\cal L}_{H \beta}(\tau)
\label{for110}
\eequ where  $h_{X_{\lambda}}$ is derived from  Stasinska's tables for  the
temperature of the equivalent star and interpolating in luminosity.

The integrated luminosity of the $X_{\lambda}$ metallic line, as sum of  
the contributions of all the HII regions, is given by:

\equ
L_{X_{\lambda}}={\dis \int_{0}^{{\tau}_{M}} } \frac{d n_{cl}}{dt} 
{\cal L}_{X_{\lambda}}(\tau) d \tau
\label{for111}
\eequ
where  $d n_{cl}/dt$ is the number of clusters of given mass and age
$\tau$ born in unit time.

\subsection{The ionized gas emission: the continuum spectrum}

The ionized gas has also a continuum emission chiefly arising from H and He.
Let us at first consider the case of a single star. The continuous
monochromatic luminosity per unit wavelength interval is obtained by: 

\equ
L_{c}(\lambda)={\dis \int_{V}} N(H^{+}) N_{e} \gamma_{T}^{c}(\lambda) dV=
N(H^{+}) N_{e} \gamma_{T}^{c}(\lambda) V
\label{for120}
\eequ

where $\gamma_{T}^{c}(\lambda)$ (Aller, 1984) includes the ratio  $\chi$ of the
radii of the ionization zones of He and H, accounting for the fact that the
region where He is completely ionized once (the temperature of all the MS
stars and WR stars is not so high to ionize He twice) does not necessarily
coincide with the region where H is completely ionized. The behaviour of
$\chi$ with the temperature of the exciting star is given by Osterbrock (1989). 

$L_{c}(\lambda)$ depends on the volume of the ionized region, which can be
eliminated with the relation: 

\equ 
n_{c}=N(H^{+}) N_{e} \alpha_{B} V
\label{for121}
\eequ
where the homogeneity of the HII region has been assumed; $n_{c}$ is the 
UV flux. Relation (\ref{for121}) is derived from the ionization balance. One
then gets: 
\equ
L_{c}(\lambda)= \frac{\gamma_{T}^{c}}{\alpha_{B}} n_{c}
\label{for122}
\eequ

Since we have approximated the integrated spectrum of each cluster with the
spectrum of an equivalent star, the continuous monochromatic luminosity of the
region powered by a cluster is given by: 

\equ 
{\cal L}_{c}(\lambda, \tau) = \frac{\gamma_{T}^{c}(\lambda)}
{\alpha_{B}}{\cal N}^{*}_{c}(\tau) 
\label{for126}
\eequ 

where ${\cal N}^{*}_{c}$ is the ionizing flux  of the equivalent star; the 
factor $\chi$ is function of the temperature of the equivalent star and is
evaluated from Fig. 2.5 of Osterbrock (1989). Table 1 shows the calibration as
function of the age for a star cluster with solar composition and an 
initial mass $M_{cl}= 3000 \,M_{\odot}$. 

The continuous monochromatic luminosity of the galaxy is derived by summing up
the contributions of all the HII regions: 

\equ
L_{c} (\lambda)={\dis \int_{0}^{{\tau}_{M}} } \frac{d n_{cl}}{dt}{\cal}L_{c}
( \lambda,\tau) 
d \tau
\label{for127}
\eequ

$d n_{cl}/dt$ being the number of star clusters born per unit time at $t$.

The results obtained by approximating the integrated spectrum of the cluster
with the spectrum of a star with suitable temperature have been compared with
models derived with the photoionization program Cloudy (Ferland, 1991) in 
which the integrated spectrum of the cluster has been used: the differences are
below the uncertainties connected with the assumptions and with the precision
with which the input physics is computed. The program Cloudy has been also used
to supplement the calibration at low temperatures ($T_{e}=19000 \, K$). 
 
The results depend obviously on the parameters taken to characterize
the clusters: a) the IMF and the upper limit of the stellar masses; throughout
this paper a Salpeter IMF with x=-1.35 and masses in the range 0.01-100 
$M_{\odot}$ have been adopted; b) the 
total number of ionizing stars in the cluster (this parameter can be changed 
in order to treat normal open clusters or more massive groups like globular 
or globular-like clusters); c) the chemical composition and the density of the 
gas. Generally the solar composition has been adopted with a temperature of
8000 K for the ionized gas while its density is taken to be $n=10$ $cm^{-3}$. 
A simple calculation of the dynamical evolution of an HII region shows that
the electron density evolves from the initial value corresponding to the
original density of the molecular gas (n=$10^3 {cm}^{-3}$) 
by two orders of magnitudes. Moreover from the ratio
of the SII lines in HII regions in spiral galaxies Zaritsky et al. (1994)
find densities ``smaller than about 100 ${cm}^{-3}$''. 
Converting Zaritsky et al. line ratios to densities according to Czyzak 
et al. (1986), densities typically less than 50 $cm^{-3}$
are found.

\section{Models and comparison with Kennicutt's spectra}

The two components of the spectrum, the stellar and gaseous components, are
correlated in the sense that the ionizing flux must be computed with the same
number of young hot stars which enter in the computation of the stellar
spectrum. To this aim the absolute visual magnitude of the galaxy has been a
priori fixed; this corresponds to determine the value of the product of the
normalization constant in the IMF and of the value of the initial SFR
$\psi_{o}$. 

Besides the first eight lines of the Balmer series and 
the $L_{\alpha}$ line at 1215
\AA, the following lines have been synthesized:
HeI5876, CII2326, CIII1909, CIV1549, NII5755, NII6584, OI6300, OII3727,
OII7325, OIII4363, OIII5007, NeIII3869, MgII2798, SII4070, SII6720, SII6312.

To determine the equivalent widths the ratio of the line flux to the total
(stars + gas) flux at the central wavelength ${\lambda}_0$ of the line is
computed. The fluxes of the H lines, given by eq. (7), are 
distributed over a gaussian
profile: 

\equ
I=\frac{I_0}{\sqrt{2 \pi} \sigma}exp{\frac{-(\lambda-{\lambda}_0)^2}
{2 {\sigma}^2}}
\eequ
$\sigma$ is such that the width of the absorption line is larger than that 
of the emission component. The equivalent width is then computed with a 
semi-automatic procedure using the program SPLOT in IRAF. 


Models have been derived to describe galaxies of several morphological types
(elliptical, Sa, Sb, Sc, Sd); the model labelled ``Extreme'' has a SFR 
monotonically
increasing along all its evolution and its metallicity evolves approximately as
in the Sd model. A further model has been derived with a SFR typical of an
elliptical, however with a smaller average metal content (Emp,
Z=0.001): this model
allows to study the influence on colour indices and spectral features of the
metallicity and corresponds to a very extreme case since ellipticals with such
a low metal content should be very faint.

The SFR adopted for normal galaxies of the morphological sequence have been
derived with a simple model of chemical evolution constrained with the broad
band  UBVRI colours, the present average metal abundance and the fraction of
gaseous mass to the total mass. The SFRs thus obtained are very similar to
those derived by Sandage (1986) and are given in the Appendix. 
An age of 15 Gyr has been adopted. 

The spectrum of the nebular component has been computed according to the 
prescriptions of sect. 3 with clusters of solar metallicity and an initial
mass of 3000 M$_{\odot}$.

Some spectral features and colours of such models are presented in Table 2 and
Table 3. The uncertainty affecting the equivalent width of the $H\beta$ line is
larger than for other lines, due the difficulty of estimating the
level of the continuum. Corresponding to maximum and mimimum estimate of it,
two extreme value have been obtained for each model and their average is shown
in Table 2. The inspection of the tables allows the estimate of the
relative importance of the two components (stellar in absorption and nebular in
emission) of the $H\alpha$, $H\beta$ and $H\delta$ lines. The models also show
that the colours of normal spirals are not significantly affected nor by the
nebular continuum nor by the emission lines. 
In fact
the contribution of the gaseous continuous
emission in spiral models
is by more than two orders of magnitude smaller than the stellar
contribution. Only in the case of very strong bursts both components are
comparable. 


A comparison with HII region models (Olofsson 1989, Garcia-Vargas et al. 1995,
Mayya 1995, Stasinska \& Leitherer 1996  and references therein) is not
straightforward, since in this work we \sl make use \rm of HII region models
(Stasinska 1990) in order to study the integrated spectrum of galaxies.

In the following analysis we compare our results with observed galaxy spectra,
considering a subsample of Kennicutt  (1992a) atlas
consisting of 33 high resolution (5 - 8 \AA) spectra and 6 medium resolution
(15 - 25 \AA) spectra covering the 3500-7500 \AA $\,$interval. This sample
includes galaxies of all the normal types. 
AGNs, in which the interstellar gas is, at least partly, ionized
by a nonthermal component, are not considered. 

A few samples of
integrated spectra of nearby galaxies, including the contribution of all
the galaxy, are available. There are in the literature 
many spectra of nuclear
regions but they cannot be used to describe the behaviour of the whole galaxy,
due to gradients of the metallicity of the gas and stellar populations.
Moreover they cannot be compared with the spectra of distant galaxies, whose
projected aperture diameter is typically of several kpc. The consequence of
this undersampling is the underestimate of the emission lines, since in several
galaxies the stellar continuum emission (originating from the disk and the
bulge) is more centrally condensed than the disk nebular emission. 

From the spectra of his sample Kennicutt (1992a, 1992b) has derived the
following quantities: the morphological type, the absolute magnitude M(B), the
equivalent width of the $H\alpha$ + [NII]$\lambda \lambda$6548-6583 blend, a
continuum colour index "41-50", the [NII]/$H\alpha$ flux ratio, the equivalent
widths and the flux ratios related to  $H\alpha$ + [NII]
of the following lines: [OII]$\lambda$3727, $H\beta$,
 [OIII]$\lambda$5007, [SII]$\lambda
\lambda$6717,6731. Uncertainties affecting such data range from 5 to 10 \% for
the galaxies with the strongest emission lines to 30 \% for galaxies with
the faintest lines. We have implemented this panorama of data by computing from
Kennicutt's spectra the quantity D$_{4000}$ and the equivalent width of the
$H\delta$ line for all the considered objects and the equivalent widths of the
[OII] and $H\beta$ lines of E and S0 galaxies, when possible. 
Furthermore the equivalent width
of the $H\alpha$ line has been derived from Kennicutt's $H\alpha$ + [NII]
equivalent width and from [NII]/$H\alpha$  flux ratios. 

The equivalent widths have been computed with the same method used for the 
synthetic spectra, the SPLOT program in IRAF. The level of the continuum and the
wavelength interval have been chosen interactively each time and in the most
difficult cases the measure has been repeated. When necessary the redshift has
been determined through the identification of the OII line at $\lambda$= 3727
\AA $\,$ and of the $H\alpha$ line in the case of emission line galaxies; in
early type galaxies the redshift value derived by Kennicutt has been adopted.
The equivalent widths thus determined
are affected by uncertainties similar to those given by Kennicutt: in the case
of the $H\delta$ line the uncertainty is of the order of 15 \%, for strong
lines it tends to be smaller and can reach also 30 \% if an emission component
is present. 

In Tables 4 and 5 all these quantities are collected for normal ellipticals and
spirals and for irregular and peculiar galaxies respectively. 

In  Fig. 1a the spectrum of the galaxy  NGC 3379 belonging to Kennicutt's
sample is compared with the elliptical model. Both emission lines and strong 
Balmer absorption lines are absent. The  Sa galaxy NGC 2775 together
with the correspondig model is shown in Fig. 1b: the  $H\alpha$ line in 
emission, although faint, is present, while $D_{4000}$ is decreased in 
comparison to the elliptical: anyway  this feature can be hardly appreciated 
by a simple eye inspection of the spectrum. The changes are more evident when
a Sc galaxy is considered as is the case of Fig. 1c where NGC 2903 is 
presented with its model: the $H\alpha$ and $H\beta$  in emission are
evident. In the observed spectrum also  the [OII]$\lambda$3727 is present 
while in the model the metallic lines are not shown although they are computed
and their equivalent widhts are presented in Table 2.

\section{Spectroscopic features affected by young stars}

In this section we analyze the behaviour of the spectral features which are
directly or indirectly connected with the presence of young stars as the Balmer
lines in emission or the [OII] line  at $\lambda=3727$ \AA. The maximum and the
average values of the equivalent width of the [OII] line increases towards late
galactic types and the spread within each morphological type is considerably
large, as shown by Kennicutt (1992b).  The trend with the type is also
confirmed by the models. 
Peculiar galaxies (mergers and starbursts) cover a wide range of values but
they have predominantly the largest equivalent widths ($EW \, >\, 40 $ \AA),
not reached by normal galaxies. 

The correlation between equivalent widths of [OII] and $H\beta$ lines is
presented in Fig. 2: within the errors there is agreement between the
theoretical results and the observations. 
Galaxies Mk 59 and Mk 71 are not reproduced in the figure: these objects, which
share the features of HII regions, deviate from the relation and have a high
EW($H\beta$)/EW([OII]) ratio ($> 1.8$) compared with the typical values ($ <
1$)  found for spirals and irregulars. 

A further test is provided by the sample of 21 nearby blue cluster or field
galaxies, mainly Sb and Sc collected by Dressler and Gunn (1982). Also in this
case spectra cover the whole galaxy ($\simeq$ 15 kpc). Again the observed
$H\beta$ - [OII] relation is reproduced with good accuracy by the models. 

This strong correlation is confirmed, for normal spiral galaxies, by the models
of Stasinska \& Leitherer (1996): they found that, if the number of ionizing
stars of an HII region is relatively small ( $\approx$ 50), the $H\beta$/[OII]
ratio is only marginally dependent on age (it is increasing with the age of 
the starburst, levelling off after a certain time), while it does depend
on the electron density. Since the equivalent width of the $H\beta$  emission
line correlates with the SFR, also EW([OII]) is a good SFR indicator. However
Stasinska \& Leitherer (1996) found that this behaviour fails when the number
of ionizing stars is considerably increased. Therefore the SFR--[OII]
calibration  is applicable to late--type galaxies, where the mass range of star
clusters and  the electron density range (Zaritsky et al. 1994) are apparently
limited. On the contrary, even in conditions typical of spiral galaxies the
$H\beta$/[OIII] ratio is strongly affected by the excitation parameters.

For a given [OII] equivalent width, models tend to have EW($H\alpha$) smaller
than observed, anyway differences are comparable to the uncertainty in the
evaluation of the equivalent width, considering the variation of the ratio
$H\alpha$/[OII] with the metallicity and the difficulty in measuring the
$H\alpha$ line due to the NII blend in the observed spectra. 

Other emission lines have been computed. In Fig. 3a and Fig. 3b the observed
relations of [SII]6716,6731  with H$_{\alpha}$ and (U-B) are shown and both are
nicely reproduced by the models. 
The [OIII] line is instead clearly overestimated in the models. In this case
the disagreement cannot be attributed to the procedure of measuring the
equivalent width from the spectra. 
This line drastically depends on the excitation conditions within the HII
regions, known to present large variations within each galaxy. This is the
reason why the [OIII] line is not fit for determining the SFR. The fact that
models systematically yield too large [OIII] values seems to suggest that the
effective temperatures adopted for the stars responsible for the ionization of
the interstellar gas may have been overestimated. This can be consequence both
of the way in which the equivalent stars are determined for the clusters
powering the HII regions and the inadequacy of Kurucz models for very hot stars
for $\lambda <$ 1000 \AA. Another possibility is the need of adopting a
different metal content in computing the gaseous emission. In fact the measured
abundance properties in nearby spiral galaxies (Zaritsky et al. 1994) seem to
indicate an oxygen content higher than solar. An inspection of Stasinska's
tables shows that when the solar metallicity is increased by a factor 2, the
ratio OIII/H$\beta$ decreases by a factor 8. 
 
Another source of uncertainty are the parameters of the IMF. While the slope
of the IMF seems not crucial in determining the stellar spectrum in young 
stellar populations (Barbaro \& Olivi 1991), a larger influence arises from
the upper mass limit. Decreasing such a limit the average temperature
of the stars responsible for the excitation of the HII region is lowered and
again the ratio OIII/H$\beta$ is significantly reduced, as it can be derived 
from Stasinska's models. Unfortunately there is no firm evidence of the 
most appropriate values of such a parameter and its dependence on the
environmental conditions.

Another quantity affecting the line ratios is the electron density but
an inspection
of Stasinska's models shows that in order to reduce the OIII/H$\beta$ ratio
smaller unrealistic values of the density are required.

\section{Spectroscopic features affected by relatively old stars}

The spectral features till now considered are related to the presence
of stars born in the last few ten million years. In the following we consider
other quantities which are the signature of stars born in previous times. 

The break at 4000 \AA $\,$ D$_{4000}$ has been often used as age indicator of
distant galaxies neglecting its dependence  on the metallicity. In Fig. 4
D$_{4000}$ is plotted against the equivalent width of the H$_{\alpha}$ line
both for normal galaxies and peculiar ones. The normal galaxies define a 
tight relation while peculiar objects
exhibit a large spread. This behaviour is expected since H$_{\alpha}$ is
affected only by the very young stars while D$_{4000}$ is influenced also by
older stars. A consequence of these consideration is that the most intense
bursts in peculiar galaxies of Kennicutt's sample must involve a large number
of young stars. They in fact have very small values of D$_{4000}$, not reached
by any normal galaxy however late it is. Nevertheless D$_{4000}$, as other 
spectral features, is able to discriminate between normal and
peculiar galaxies only in the most extreme cases. 

The D$_{4000}$ - OII and D$_{4000}$ - H$_{\beta}$ relations are very similar
and lead to the same conclusions: D$_{4000}$ is a good age indicator only in
the case of normal galaxies, but fails in the case of starbursts.


Unlike the Balmer lines up to now considered, the $H\delta$ line is always
observed in absorption in the spectra of spiral galaxies, since the
stellar absorption overwhelmes the gaseous emission. Most of
the normal galaxies of the sample have EW($H\delta$) $<$ 3.2
\AA. Only two Sc spirals have 3.2 $<$ EW($H\delta$) $<$ 4. All the peculiar
objects, with the exception of the "E+A" galaxy, have EW($H\delta$) $<$ 2 and
in the spectra of many of them the line is emission-filled. Based on our
models, emission $H\delta$
lines are expected only in the spectra of objects with a current intense
starburst.

Strong absorption $H\delta$ lines instead are typical of galaxies which have
experienced in the past an intense burst of star formation ended about 1 - 2
gyr ago. Their spectra can be interpreted as the superposition of a normal
galaxy (elliptical or spiral) and a generation of A stars (Dressler \& Gunn 
1982).
A detailed analysis (Poggianti and Barbaro, 1996) dealing with all stars having
enhanced $H\delta$ absorption confirms the validity of this interpretation. In
the adopted sample this is the case of NGC 3921, which due to the strong
absorption Balmer line has been classified as E+A. 
This interacting galaxy is
probably in an advanced burst phase or even in a post-burst phase in which
the emission lines are weak and the absorption feature begin to emerge. 


The equivalent widths of the $H\alpha$ and [OII] lines are well correlated
with (U-B) e (B-V) colours (Fig. 5a and Fig. 5b) and these relations are nicely
fitted by the models of normal galaxies (in such diagrams E-Sa-Sb-Sc-Sd models
are shown). The correlations between the colours and the equivalent widths
become weaker in the case of peculiar galaxies; these objects show an excess of
emission with respect to what is expected by extrapolating the relation found for
normal galaxies. 
The information emerging from the (U-B) e (B-V) colours is therefore
equivalent to that supplied from the spectral features in the case of a galaxy
whose star formation history did not show sudden changes. 

We have seen that a net flux of 
the $H\delta$ line in emission testifies the presence of a
current burst while a strong absorption feature is indicative of a burst ended
some 1 - 2 Gyr ago. Some starbursts can be also
identified by their extremely blue colours or very low $D_{4000}$ values
(see Tables 2 and 3 for the lowest values of colours and $D_{4000}$ for normal
galaxies, given by the ``Extreme'' case). 

\section{Evaluation of the star formation rate}

In this section an attempt is made to exploit the general features of the
models and the peculiar characteristics of some quantities to derive estimates
of the star formation rates in normal and starburst galaxies. A first
evaluation, limited to the normal galaxies of the Hubble sequence, can be
obtained from purely photometric data. To this aim spectrophotometric models
have been derived by using an analytical SFR of the type: 

\equ
\psi(t)=\psi_{o} \beta exp(- \beta t)
\label{for888}
\eequ

This relation is a fairly good approximation of the SFR so far adopted and
represents a suitable parametrization. The $\beta$ parameter is related to the
timescale of the star formation, $\psi_{o}$ depends on the galaxy mass. The UBV
colours depend solely on the relative distribution of stars of different age
and therefore are function of $\beta$ alone. With a SFR of this type and with
convenient values of $\beta$ the colours of the different morphological types
are satisfactory reproduced. 

The (U-B) colour index as a function of $\beta$ is shown in Fig. 6. Provided
that the correction for reddening is determined it is an easy matter to derive
with this relation the value of $\beta$. To ignore the reddening correction
implies an overestimate of the $\beta$ value. 


The luminosities in the different photometric bands are instead function both
of $\beta$ and of the galaxy mass and therefore of $\psi_{o}$. In Fig. 7 the 
$M_V$ magnitude is plotted against $\beta$ for several values of $\psi_{o}$: 
entering this diagram with $M_V$ and $\beta$, the value of $\psi_{o}$
can be determined and consequently both parameters of $\psi(t)$ are estimated.

A further calibration can be derived for late-type galaxies in terms of the
$D_{4000}$ index. We still use the expression (\ref{for888}) for $\psi(t)$.
For each value of $\beta$ the ratio $<\psi>/\psi(T)$ between the average value
of $\psi$ over the time interval (T- $\delta t$,T) and the present value of 
$\psi$ can be derived. The average value is given by:

\equ
<\psi>=\frac{1}{\delta t} \dis{\int_{T -\delta t} ^{T}} \psi_{o} \beta
exp(-\beta t) dt
\eequ
\equ
=\frac{\psi_{o} \beta}{\delta t} \frac{exp(- \beta (T-\delta t)
-exp(- \beta T)}{\beta}
\eequ

One then gets:

\equ
\frac{<\psi>}{\psi(T)}= \frac{exp(\beta \delta t)-1}{\beta \delta t}
\label{for812}
\eequ

This quantity can be easily computed as a function of $\beta$ for several values
of $\delta t$: only for large values of  $\delta t$ does the function
$<\psi>/\psi(T)$ vary over a sufficient large interval allowing an accurate
evaluation. Therefore we adopt $\delta t= 1$, 5 and 16 Gyr.
From (\ref{for812}) and from the D$_{4000}$ - $\beta$ relation derived
from the models the calibration $<\psi>/\psi(T)- D_{4000}$, shown in Table 6, 
is
obtained. From the observed value of D$_{4000}$ the ratio $<\psi>/\psi(T)$ is
then estimated and, if $\psi(T)$ is derived from one method described in the
following, the average SFR over the last billion years can be evaluated. Since
$\psi(T)$ can be determined only for late-type galaxies the method is
restricted to object with $D_{4000} < 2.1$. The same calibration is obtained 
using a non analytical SFR, as in the majority of the models considered in 
this work. Note that $<\psi>/\psi(T)$ for  $\delta t= 16 $ Gyr is the inverse 
of the Scalo's (1986) b parameter.

In particular eq. (\ref{for888}) allows the determination of the present SFR. This
quantity can be estimated also from the spectral features, in particular from
the emission nebular lines. In ionization-bound  HII regions the luminosity of
the Balmer emission lines is, to a first approximation, independent of the
temperature of the gas and is only related to the total ionizing flux of hot
young stars (Osterbrock 1989). This is the theoretical reason for deriving the
formation rate of massive stars from the Balmer line fluxes.  When an estimate
of the IMF is made, the total SFR can be also derived. A series of models of
spirals of different morphological types with M$_{B}=-18$ mag and M$_{B}=-20$
mag has been computed in order to derive the relations between the SFR (in
$M_{\odot} \, yr^{-1}$) and the luminosities of the $H\alpha$ and OII lines (in
$ergs \, s^{-1}$) and their equivalent widths.  

Using  spiral models later than Sa one has for the luminosity of the $H\alpha$
line: 

\equ
\psi = \, 4 \, \times \, 10^{-41} L(H\alpha)
\label{for100}
\eequ 

This calibration can be compared with a similar relation found by Kennicutt
(1983) in which the numerical coefficient is $ 2.3 \, \times \, 10^{-41}$. The
difference is partly due to different IMFs used in the two calibrations.
Kennicutt (1983) argues that the H$\alpha$ line is the best feature for
estimating the SFR and he points out
that for distant galaxies, due to redshift, the
H$\alpha$ line falls outside the visible spectrum and in its place the [OII]
line is usually employed. From the set of our models the following
relation has been derived: 

\equ
\psi = \, 6.3 \, \times \, 10^{-41} L(OII)
\label{for211}
\eequ 

Since the [OII] line flux is related to the total number of massive stars able
to ionize the gas and since the maximum main sequence lifetime of these stars
is some ten million years, $\psi $ is the 
SFR averaged over the last  $10^7$ yr. This
relation is universal in the sense that it does not depend on the overall
history of the star formation and therefore it applies both to normal galaxies
and to starbursts and, provided that there has been no abrupt change, can be
thought as the present SFR. Of course if the burst started less than  $10^7$ yr
the SFR cannot be derived from the [OII] flux unless the length of the burst
can be guessed. 

Starting from his calibration equivalent to (\ref{for100}) and using a relation
between the H$\alpha$ and [OII] fluxes, derived from his sample, Kennicutt
(1992b) obtained an analogous relation for the luminosity of the [OII] line : 

\equ
\psi(M_\odot yr^{-1}) \, = \, 5.0 \, \times \, 10^{-41} L([OII])
\label{for14}
\eequ to be compared with our equation (\ref{for211}). The  differences between
(\ref{for211}) and (\ref{for14}) are of the same order of magnitude as in the
L($H\alpha$) -- $\psi$  relations. The uncertainty affecting eq. (\ref{for14})
are more important than that of L($H\alpha$) -- $\psi$  relation because of the
large spread affecting the relations from which  has been derived. 

A different calibration of the SFR in terms of the [OII] luminosity has been
obtained by Gallagher et al. (1989) by measuring the [OII] and H$\beta$ fluxes
of a variety of blue objects of different types (irregular, amorphous,
unclassified galaxies) with a pronounced star formation activity and a
relatively low reddening. The numerical constant in their relation is lower
than that of Kennicutt's calibration by a factor of 5. It is hard to ascertain
to what extent such differences are related to differences in the two samples
and  to what extent they depend on the adopted approximations. The comparison
of Kennicutt's results with those of Gallagher allows to estimate an
uncertainty of a factor of 2 or 3 in the evaluation of the SFR. 

Line fluxes however are not easily achieved for distant galaxies, for which
instead the equivalent widths are measured. It is therefore important to
inquire how such quantities can be used to estimate the present SFR. The
equivalent width is approximately equal to the ratio of the line flux to the
continuum flux at the central wavelength of the line. In the case of the [OII]
line we have: 

\equ
EW(OII) \simeq \frac {L(OII)}{F_{\lambda =3727}}
\label{for881}
\eequ

and therefore EW(OII) is directly proportional to the present SFR but is also
depends on the continuum flux at the central wavelength of the line, which
generally is built up with the contribution of all the stars and, which, to a
first approximation, is proportional to the luminosity of the nearest
photometric band. In the case of the [OII] line this is the U band of the
Johnson system. Both $L(OII)$ and $F_{\lambda =3727}$  depend in the same way
on the mass of the galaxy and therefore the equivalent width is the same for
two galaxies with the same shape of the SFR but different masses. This
behaviour is confirmed by a series of models with a wide range of ages and
different star formation histories. These models reproduce both normal local
morphological types and bursts of different intensity and length in ellipticals
and spirals (Sa and Sc). 

In Fig. 8, where the equivalent width and the luminosity of the OII line are
reproduced, three models are shown with $EW=46$ \AA $\,$ and with completely
different star formation histories: one point corresponds to an elliptical with
burst and the the other two to spirals with different bursts.
 Models lying on
the dashed line on the top of the diagram instead have SFRs with  the same
shape but  differ by the value of M$_{B}$ which ranges from -18 to -21 mag.
Such large spread appears also from the analogous observed diagram of Gallagher
et al (1989).  It is then confirmed that only the shape of the SFR affects the
equivalent width while the line luminosity depends both on its shape and the
mass of the galaxy. 

A relation between flux and equivalent width has been 
obtained directly from the models:

\equ
\psi = 1.1 \times L_{U} [EW(OII) + 1.46]
\label{for871}
\eequ

where L$_{U}$ is in units of $10^{42}$ ergs  per second. This relation allows 
one to derive the mean SFR averaged over the last 10$^{7}$ yr from EW(OII)
provided that $L_{U}$ is also known.

From a subsample of 16 galaxies Kennicutt (1992b) derived a  mean relation
between the  $L_B$ luminosity and the equivalent width of the [OII]  line: 

\equ
L(OII) \simeq 1.4 \times 10^{29} \cdot L_B \cdot EW(OII) ergs \, s^{-1}
\label{for112}
\eequ
where $L_B$ is given in units of L$_{B, \odot}$. 

This equation is the analogous of equation (\ref{for881}) with the difference
that $L_B$ is used instead of $L_{U}$. Both relations would have the same field
of applicability if $L_B$ and $L_{U}$  were proportional. A simple relation 
between the two luminosities
exists only for normal morphological types, while it
breaks down in the case of starbursts. Therefore equation (\ref{for112}) seems
to be less general than (\ref{for881}). When a burst is present $L_{U}$ does
not depend in a simple way on the mean SFR over some time period. 

The methods of evaluation the star formation rates described in this section
have been applied to the normal galaxies of Kennicutt's sample and the results
are collected in Table 7.

The values of $\beta$ have been computed from the observed (U-B) colour index
with the calibration of Fig. 6. The mean value of $\beta$ for each
morphological type progressively decreases starting from  the ellipticals
($1$ $Gyr^{-1}$),  S0 ($0.43$), Sa, Sab ($0.28$), Sb, Sbc 
($0.1$ ), Sc ($0.055$). This result is in
agreement with the well-established interpretation of the colour indices and
the spectroscopic properties of galaxies along the Hubble sequence and provides
the same conclusions reached by Kennicutt et al. (1994) with the analysis of
Scalo's b index. Table 7 clearly shows the large spread in $\beta$
within each Hubble type thus confirming that a close correspondence does not
exist between "morphological" and "spectrophotometric" types, as colour indices
and spectral features already have shown. For each galaxy from the observed
blue magnitude and from the value of $\beta$ previously derived the values of
$\psi_{o}$ have been computed with the use of Fig. 7. The SFR computed with
(\ref{for888}) at the age of 16 Gyr, in M$_{\odot}$/yr, is shown in column
(10), while in column (11) the evaluation of the 
same quantity is given as derived
from the observed [OII] equivalent widths according to (\ref{for871}). The
difference in the results obtained with the two methods is of the same order of
magnitude as the uncertainties affecting both procedures. 
The present evaluation of Scalo's b parameter is also shown in column (12).

Kennicutt et al. (1994) also performed an analysis with the aid of a
spectrophotometric model of EW(H$\alpha$) and some colours of nearby spiral
galaxies but their results cannot be compared with the 
present outcomes due to the different assumptions concerning the age of 
the galaxies.

With the previous considerations relations have been derived able to estimate
the SFR from the H$\alpha$ and [OII] luminosities for every type of galaxies,
normal or starburst. Other methods are restricted only to normal galaxies. It
is therefore urgent to derive criteria to diagnose the presence or the absence
of bursts. It has been pointed out that very large values of the equivalent
width of the H$\alpha$ line (EW $>$ 50 \AA$\,$) and the OII line (EW $> 50$
\AA $\,$, if also lower than solar metallicities are considered) are
typical of starbursts. Also negative H$\delta$ equivalent widths are indicative
of a bursts since in late type systems this line is at most emission-filled. On
the contrary positive values of EW(H$\delta$) larger than 3 are indicative
of a galaxy which hosted a burst ended 500 - 1000 Myr ago. 
Therefore the  EW(OII)- EW(H$\delta$) diagram is a valid tool to single 
out starbursts. With
reference to the Fig. 9 normal galaxies are located in a central band with $0
\ge EW(H\delta) \le 3$ and $EW(OII) \leq 50$ \AA. Models with a current burst
fall on the left/top of this band; the more intense and of smaller length is
the burst, the farthest is its location from the band. Models of the same
intensity with increasing length tend to approach the more to the band: this
behaviour is selfexplanatory  since both line fluxes become saturated, being
connected to the number of stars younger than some 10$^{7}$ years while the
continuum keeps increasing. A consequence of this is that weak bursts 
and those intense and of large length can get confused with normal galaxies.
In Fig. 9 for a current burst the line $\psi_{1}/\psi_{o} = const$ is drawn. 

Models in the post-burst phase fall in the region on the right of the band of
normal systems; the farther they are from it the larger was the number of stars
produced by the burst. These post-starburst models assume the star formation 
stops at the end of the burst and therefore have EW(OII)=0.

Such lines are independent of the type of underlying galaxy, provided that the
burst is sufficiently intense: in such situations an estimate of the intensity
of the burst is possible. In the diagram the objects of Kennicutt's sample are
shown: they confirm the conclusions reached from the analysis of the models.
Obviously the same information is provided by the EW(H$\alpha$)-EW(H$\delta$)
diagram.

\section{Conclusions}

The comparison with Kennicutt's spectra has shown that the model nicely
reproduces the emission lines of observed spectra, with the exceptions 
of the OIII line at $\lambda= 5007$, as previously discussed. 

To a first approximation the relations between the intensities of the 
emission lines are similar both for normal galaxies and those showing a
stronger astrogenetic activity: this is reasonable since all such quantities
are only affected by young stars. However peculiar objects present a larger
spread, consequence of the variety of the parameters characterizing the bursts.

The analysis performed with the models confirms and extends Kennicutt's
conclusions (1992b): the $H\alpha$ and [OII] luminosities are good indicators
of the present star formation. 
The equivalent widths of these lines, which usually are the only
available data in the case of distant galaxies, can be also profitably used
together with the continuum flux at the line wavelength. This last quantity can
be estimated by means of a nearby photometric band: in the case of the [OII]
line the nearest one is the U band. 

The occurrence of current bursts can be ascertained by means of exceptionally
large equivalent widths of  H$_{\alpha}$, [OII]  emission lines, by the 
presence of the H$\delta$ line in emission and by very small values
of the D$_{4000}$ index. A large equivalent width (EW $ > 3 \AA$) of the
H$\delta$ line in absorption is instead indicative of a burst already ended (by
about 1-2 Gyr). It must be pointed out however that bursts cannot always be
singled out. For instance models of galaxies with a burst with particular
values of the parameters can fall in the  EW(OII) - EW(H$\delta)$ diagram just
in the region filled by normal objects and this happens not only with the weak
and short lasting bursts. 

The information supported by the (U-B) e (B-V) colours is equivalent to that
arising from the emission  lines only in the case of normal galaxies. Very blue
colour indices also witness current bursts. In the case of normal galaxies the
history of star formation can be estimated from photometric data. The (U-B), or
another broad band colour index, can be used to evaluate the e-folding time of
the star formation process, while $M_V$, for instance, gives then the initial
value of the SFR. The D$_{4000}$ index has been calibrated to yield the ratio
of the SFR averaged over the last 5 billion years to the present SFR. If this
last quantity is derived from the emission lines, the mean SFR can be
estimated. 

Attempts to give quantitative estimates of the SF histories in galaxies
hosting a burst generally fail, essentially because of the large number of
parameters involved.

\begin{acknowledgements} 
We acknowledge the
availability of the Kennicutt's galaxy atlas and the Jacoby et al's stellar
library from the NDSS-DCA Astronomical Data Center.
B.M.P. acknowledges the receipt of a grant from the Department of Physics
of the University of Pisa.
This work was supported in part by the Formation and Evolution of Galaxies
network set up by the European Commission under contract ERB FMRX-CT96-086 of 
its TMR programme.
\end{acknowledgements}

\appendix
\section{Appendix}

The SFRs of the different Hubble types have been obtained by constraining the 
spectrophotometric galaxy  models with the present day gas fraction, metal 
content and UBV colours. This work is part of an unpublished Master Degree
Thesis (Auddino, University of Padova 1992).

The results for spiral galaxies are shown in Table 8.
The first column gives the age in years while the others show the
SFR in arbitrary units for each type. 
Table 9 gives the resulting average and present
metallicity, the present SFR (in $M_{\odot} yr^{-1}$ if the galactic mass is
$10^{11} M_{\odot}$) and fraction of gaseous mass for an age of 16 Gyr.

\clearpage

\bf
\centerline{Figure captions}
\rm
\noindent
{\bf Fig. 1a-c. }Comparison between observed and synthetic spectra, in 
arbitrary flux units:
\bf a \rm NGC3379 (top) and elliptical model (bottom), 
\bf b \rm NGC2775 (top) and Sa model (bottom), \bf c \rm NGC2903 (top)
and Sc model (bottom) \medskip

\noindent
{\bf Fig. 2. }Equivalent widths of [OII] and $H\beta$ of the galaxies
of Kennicutt's sample (empty symbols: squares are ellipticals and spirals
from Table 4, circles are irregular and peculiar galaxies from Table 5)
and models (filled symbols). The error bars indicate the
uncertainties in the placement of the continuum 
\medskip

\noindent
{\bf Fig. 3a-b. }[SII] equivalent widths 
versus EW($H\alpha$) (\bf a) \rm and (U-B) colour index
(\bf b) \rm; symbols as in Fig. 2
\medskip

\noindent
{\bf Fig. 4. }$D_{4000}$ versus EW($H\alpha$); symbols as in Fig. 2
\medskip

\noindent
{\bf Fig. 5a-b. }Comparison emission line--colours; symbols as in Fig. 2
\medskip

\noindent
{\bf Fig. 6. }(U-B) color index versus $\beta$ ($Gyr^{-1}$) for models
with an analytical SFR (eq. 19)
\medskip

\noindent
{\bf Fig. 7. }$M_V$ magnitude as a function of $\beta$ for different values
of ${\psi}_0$
\medskip

\noindent
{\bf Fig. 8. }Fluxes and equivalent widths of [OII] for a series of burst
models in ellipticals and spirals with $M_B =-20$ (open circles) with 
different burst parameters; the dashed line connects models with the same
SF history and different absolute magnitudes (filled circles, from left to
right $M_B =-18, -18.5, -19, -19.5, -20, -21$)
\medskip

\noindent
{\bf Fig. 9. }EW(OII) versus EW($H\delta$). The shaded area represents the
region of the diagram where models of the normal spectral types lie.
In such area, the highest EWs(OII) can be reached considering
less than solar metallicities. Burst models with different burst lenghts and
constant intensity are connected by the solid line: from left to right the
points indicate bursts with lenght 10, 100 and 1000 Myr.
Post-starburst models in which the star formation is truncated at the end of
the burst are also shown: crosses represent models of a burst in a Sc spiral
galaxy in which the
star formation ended 100 Myr ago and involving 0.3, 3 and 26 \%
of the galactic mass (from left to right). The dashed line shows the time 
evolution of the 26 \% model at 0,1,10,30,100 Myr after the end of the burst
(from left to right).
Kennicutt's data are presented 
as open symbols (squares are ellipticals and spirals, circles are irregulars
and peculiar galaxies). See text for details
\medskip

\clearpage


\BTA

\caption[]{Adopted calibration values for a star cluster of
different ages: temperature of the equivalent star, adopted $\chi$ parameter
and logarithm of the number of ionizing photons}

\begin{flushleft}

\TAB{rrrc}
\hline
\noalign{\smallskip}
Age & $T_{equiv.}$ & $\chi$ & log NUV \\
\noalign{\smallskip}
($10^6 yr$) &   (kelvin)     &        &  ($M_{cl}=3000 M_{\odot}$) \\
\noalign{\smallskip}
\hline
\noalign{\smallskip}
1           & 45000        &   1     &  49.213\\
3           & 45000        &   1     &  49.301\\
4           & 40000        &   1     &  48.935\\
10          & 40000        &   1     &  47.876\\
40          & 19000        &   0     &  43.739\\
\noalign{\smallskip}
\hline
\ETAB

\end{flushleft}

\ETA

\clearpage


\BTA

\caption[]{Model results as a function of the spectral type. 
A positive value of the EW(${H\delta}$) indicates the line is in     
absorption, while the opposite is true for the other lines.
Results for the three Balmer lines are shown for emission only (emis)
and including both absorption and emission (tot).  Also shown are the 
``Extreme'' model (constantly increasing SFR) and the metal poor elliptical
model (Emp) with average Z=0.001}

\tabcolsep 0.05truecm
\arrayrulewidth 0.01truecm
\doublerulesep 0.0001truecm

\begin{flushleft}

\TAB{lrrrrrrrrr}          
\hline
\noalign{\smallskip}
Type & $D_{4000}$ & EW(OII) & EW($H\delta$) & EW($H\alpha$) & EW($H\beta$) &
EW(OIII) & EW($H\delta$) & EW($H\alpha$) & EW($H\beta$)\\
\noalign{\smallskip}
& & & (tot) & (tot) & (tot) & & (emis) & (emis) & (emis) \\ 
\noalign{\smallskip}
\hline
\noalign{\smallskip}
E  &2.21 & 0.0 & 0.9&  -1.0 & -1.7 & 0.0&  -- &  -- &   -- \\   
Sa &1.91 & 4.7 & 1.1&   1.7 & -2.2 & 2.0& -0.4&  3.2&  1.2 \\
Sb &1.75 & 9.0 & 1.0&   6.0 & -0.7 & 4.7& -0.9&  7.8&  2.8 \\
Sc &1.54 &14.9 & 1.4&  16.5 &  1.7 &10.2& -1.8& 19.2&  6.3 \\
Sd &1.42 &17.3 & 1.7&  26.3 &  3.4 &14.1& -2.4& 29.6&  9.0 \\
Ex &1.31 &21.7 & 1.2&  47.4 &  6.8 &21.9& -3.3& 52.8& 14.6 \\
Emp&1.72 & 0.0 & 1.2&  -1.0 & -1.7 & 0.0&  -- &   --&   -- \\
\noalign{\smallskip}
\hline
\ETAB

\end{flushleft}

\ETA

\clearpage


\BTA

\caption[]{Model results as a function of the spectral type: colour indices}

\tabcolsep 0.05truecm
\arrayrulewidth 0.01truecm
\doublerulesep 0.0001truecm

\begin{flushleft}

\TAB{lrrrrrrrrr}          
\hline
\noalign{\smallskip}
Type & (1550-V) & (U-B) & (B-V) & (V-R) & (R-I) &
(g-r) & (r-i) & (U-685) & ($B_j$ - $R_f$)\\
\noalign{\smallskip}
\hline
\noalign{\smallskip}
E  & 5.37 &0.50&0.93&0.74&0.65&0.71&0.39&2.77&1.71\\   
Sa & 1.09 &0.31&0.83&0.68&0.57&0.64&0.33&2.40&1.56\\
Sb & 0.16 &0.19&0.76&0.65&0.56&0.60&0.32&2.19&1.47\\
Sc &-0.74 &0.00&0.63&0.58&0.52&0.52&0.29&1.80&1.29\\
Sd &-1.17 &-0.11&0.53&0.52&0.46&0.44&0.25&1.52&1.14\\
Ex &-1.72 &-0.24&0.39&0.45&0.44&0.34&0.23&1.18&0.94\\
Emp& 3.79 &0.21 &0.77&0.58&0.44&0.55&0.23&2.15&1.43\\
\noalign{\smallskip}
\hline
\ETAB

\end{flushleft}

\ETA

\clearpage


\BTA

\caption[]{Normal galaxies in Kennicutt's sample: (1) name; 
(2) morphological type; (3) an asterisk indicates a low resolution spectrum;
(4) $D_{4000}$; (5),(6),(7),(8),(9) equivalent widths in \AA; note that 
a positive value of the EW(${H\delta}$) indicates the line is in
absorption, while the opposite is true for the other lines; (10),(11) fully
corrected colours from the Third Reference Catalogue of Bright Galaxies 
(De Vaucoulers et al. 1991)}

\tabcolsep 0.05truecm
\arrayrulewidth 0.01truecm
\doublerulesep 0.0001truecm

\begin{flushleft}

\TAB{llcrrrrrrrr}          
\hline
\noalign{\smallskip}
Name & Type & Res & $D_{4000}$ & EW(OII) & EW(${H\delta}$) &
 EW(${H\alpha}$) & EW(${H\beta}$) & EW(OIII) & $(B-V)$ & $(U-B)$  \\
\noalign{\smallskip}
(1)  & (2) &  (3) &  (4) & (5) & (6) & (7) & (8) & (9) & (10) & (11) \\
\noalign{\smallskip}
\hline
\noalign{\smallskip}
NGC3379 & E0    &   & 2.23 &  0.0&  0.7&    -- & -2.0&  0.0 & 0.94&  0.52\\
NGC4472 & E1/S0 & * & 2.10 &  0.0&  0.0&  -1.0 & -1.4&  0.0 & 0.95&  0.56\\ 
NGC4648 & E3    &   & 2.22 &  0.0&  0.9&  -1.0 & -2.0&  0.0 & 0.89&  0.51\\  
NGC4889 & E4    &   & 2.26 &  0.0&  0.0&    -- &   --&  0.0 & 0.97&  0.54\\ 
NGC3245 & S0    &   & 2.09 &  0.0&  0.7&  -0.9 & -2.5&  0.0 & 0.86&  0.44\\
NGC5866 & S0    & * & 2.00 &  0.0&  0.0&    -- &   --&  0.0 & 0.79&  0.33\\
NGC4262 & SB0   &   & 2.13 &  2.0&  1.2&    -- &   --&  --  & 0.90&  0.49\\
NGC3941 & SB0/a &   & 2.17 &  1.0&  1.1&    -- &   --&  --  & 0.88&  0.43\\ 
NGC1357 & Sa    &   & 1.84 &  4.0&  1.7&   5.4 & -2.5& $<0.3$ & 0.80&  0.20\\ 
NGC2775 & Sa    &   & 2.00 &  1.6&  1.2&   1.1 & -4.0& $<0.5$ & 0.85&  0.33\\ 
NGC3623 & Sa    &   & 2.10 &  1.5&  1.4&   0.7 & -3.0& $<0.5$ & 0.81&  0.35\\
NGC3368 & Sab   &   & 1.97 &  4.0&  1.5&   1.7 & -3.5& $<0.5$ & 0.81&  0.27\\
NGC3147 & Sb    &   & 1.79 &  3.0&  1.9&   7.5 & -1.0& $<0.5$ & 0.78&    --\\ 
NGC3627 & Sb    &   & 1.60 &  5.0&  3.1&  13.7 & -3.0& $<0.4$ & 0.66&  0.14\\  
NGC1832 & SBb   &   & 1.50 & 17.0&  1.8&  23.8 &  0.0&  2.0 & 0.54& -0.08\\ 
NGC5248 & Sbc   &   & 1.52 &  9.0&  2.3&  21.8 &  1.0&  1.0 & 0.62&  0.02\\ 
NGC6217 & SBbc  &   & 1.55 & 12.0&  3.2&  26.1 &  0.0&  2.0 & 0.57& -0.22\\ 
NGC2903 & Sc    &   & 1.54 &  8.0&  3.1&  19.0 & -1.5&  $<0.6$ & 0.59&  0.00\\ 
NGC4631 & Sc    &   & 1.32 & 40.0&  3.6&  49.6 &  4.0&  15.0 & 0.41&    --\\ 
NGC6181 & Sc    &   & 1.43 & 15.0&  2.3&  33.3 &  1.5&  3.5 & 0.49& -0.13\\ 
NGC6643 & Sc    &   & 1.60 & 11.0&  3.7&  25.4 &  0.0&  1.5 & 0.52& -0.13\\ 
NGC4775 & Sc    &   & 1.29 & 35.0&  2.1&  45.9 &  6.0&  14.0 &   --&    --\\
\noalign{\smallskip}
\hline
\ETAB

\end{flushleft}

\ETA

\clearpage


\BTA

\caption[]{As Table 4 for irregulars and peculiar galaxies. Symbols
in the last column (from Kennicutt): (a) 
spectrum with high Galactic reddening; (b)
interacting/merging; (c) starburst nucleus; (d) E+A;
(e) global starburst; (f) aperture centered on the dominant
HII region in large galaxy}

\tabcolsep 0.05truecm
\arrayrulewidth 0.01truecm
\doublerulesep 0.0001truecm

\begin{flushleft}

\TAB{llcrrrrrrrrc}          
\hline
\noalign{\smallskip}
Name & Type & Res & $D_{4000}$ & EW(OII) & EW(${H\delta}$) & 
EW(${H\alpha}$) & EW(${H\beta}$) & EW(OIII) & $(B-V)$ & $(U-B)$ & Notes \\
\noalign{\smallskip}
(1)  & (2) &  (3) &  (4) & (5) & (6) & (7) & (8) & (9) & (10) & (11) & (12) \\
\noalign{\smallskip}
\hline
\noalign{\smallskip}
NGC4449 & Sm/Im &   & 1.27 & 55.0&  1.9&  75.0 & 11.0&  33.0 & 0.37& -0.38& \\
NGC1569 & Sm/Im &   & 1.40 & 49.0&  0.4& 121.7 & 35.0&  182.0 &   --&    --& (a) \\
Mrk59   & SBm/Im& * & 0.79 & 62.0& -8.2& 765.0 &115.0&  750.0 &  --&    --& (f)\\
Mrk71   & SBm   & * & 0.82 &108.0&-60.0&1323.5 &299.0&  2490.0 &   --&    --& (f)\\
NGC4485 & Sm/Im &   & 1.30 & 48.0&  2.2&  59.1 &  4.0&  33.0 & 0.35& -0.25& (b) \\
NGC2276 & Scpec &   & 1.25 & 33.0&  1.0&  64.6 &  7.0&  10.0 & 0.45& -0.14& (b) \\ 
NGC3690 & Scpec & * & 1.14 & 48.0&  0.0& 168.3 & 22.0&  30.0 & 0.59&  0.04& (b) \\ 
NGC3303 & Pec   & * & 1.89 & 35.0&  3.2&    -- & -3.0&  2.0 &   --&    --& (b) \\
NGC7714 & Spec  &   & 1.03 & 59.0& -1.7& 139.8 & 22.0&  41.0 & 0.44& -0.51& (b) (c) \\
NGC2798 & Sapec &   & 1.53 & 17.0&  2.1&  41.7 &  1.0&  3.0 & 0.63& -0.07& (b) (c) \\    
NGC3921 & S0pec &   & 1.78 & 11.0&  4.5&   8.5 & -3.0&  $<0.5$ & 0.59&  0.26& (b) (d)\\ 
NGC3310 & Sbcpec&   & 1.19 & 63.0&  0.3& 126.4 & 20.0&  48.0 & 0.32& -0.45& (e)\\ 
UGC6697 & Spec  &   & 1.16 & 44.0&  1.1&  62.6 &  8.0&  22.0 & 0.28& -0.48&\\
NGC4750 & Sbpec &   & 1.66 &  5.0&  2.0&  10.4 & -2.0&  $<0.5$ &  --&    --&\\
NGC4670 & SBpec &   & 1.32 & 62.0&  2.1&  93.4 & 22.0&  54.0 & 0.36& -0.50&\\
\noalign{\smallskip}
\hline
\ETAB

\end{flushleft}

\ETA

\clearpage


\BTA  

\caption[]{Ratio of the past average SFR to the present SFR
${<\psi>}/{\psi(T)}$ as a function of  $D_{4000}$; the 
average past SFR is obtained over three time intervals: 1, 5 and 16 Gyr}

\begin{flushleft}

\TAB{ccccc}
\hline
\noalign{\smallskip}
$\beta$  & $D_{4000}$  &  ${<\psi>}/{\psi(T)}$ &  ${<\psi>}/{\psi(T)}$ & 
${<\psi>}/{\psi(T)}$\\
\noalign{\smallskip}
($Gyr^{-1}$) &   & $\delta t = 1$ Gyr & $\delta t= 5 $ Gyr & $\delta 
t = 15$ Gyr \\
\noalign{\smallskip}
\hline
\noalign{\smallskip}
1.0  &  2.21  &  1.72  &   29.48   &   $5.55 \times 10^{5}$  \\ 
0.7  &  2.20  &  1.45  &    9.18   &   $6.53 \times 10^{3}$  \\
0.5  &  2.15  &  1.30  &    4.47   &   $3.72 \times 10^{2}$  \\
0.3  &  1.87  &  1.17  &    2.32   &   $2.51 \times 10^{1}$  \\
0.2  &  1.65  &  1.11  &    1.72   &     7.35            \\
0.1  &  1.48  &  1.05  &    1.30   &     2.47            \\
0.05 &  1.42  &  1.03  &    1.14   &     1.53            \\
0.01 &  1.38  &  1.01  &    1.03   &     1.08            \\       
\noalign{\smallskip}
\hline
\end{tabular}

\end{flushleft}

\end{table}

\clearpage

\begin{table}  

\caption[]{SFRs of the galaxies in Kennicutt's sample. Values of $\beta$ in
$Gyr^{-1}$ are derived from the observed (U-B) by using Fig. 6.
The luminosities $L_V$ and $L_U$ are derived from the models and are given in
the tables in units $10^{42} ergs \, s^{-1}$. Present SFRs are obtained with
two methods (see text for details)}

\begin{flushleft}

\begin{tabular}{llcccccccccc}
\hline
\noalign{\smallskip}
name & type  & (U-B) & beta & $M_B$ & $L_V$ & $L_U$ & EW(OII) 
&  ${\psi}_0$ & ${SFR}^1$ & ${SFR}^2$ & b  \\
\noalign{\smallskip}
(1)&(2)&(3)&(4)&(5)&(6)&(7)&(8)&(9)&(10)&(11)&(12)\\
\noalign{\smallskip}
\hline
\noalign{\smallskip}
n3379  & E0  &  0.52 &1.00 &  -19.7 & 7.12& 1.32&   0 &  53&  $6\times {10}^{-5}$ &   -- & $1.8 \times {10}^{-6}$ \\
n4472 &E1/S0 &0.56 &1.00 &  -21.5 & 37.38& 6.94&    0 &  -- &   -- &                 -- & $1.8 \times {10}^{-6}$ \\
n4648 &E3 &   0.51& 1.00 &  -19.0 & 3.74& 6.94&    0 &  30 & $3\times {10}^{-5}$ &   -- & $1.8 \times {10}^{-6}$ \\
n4889 &E4 &   0.54 &1.00 &  -22.1 & 64.96& 12.06 &  0 &  -- &   --  &                -- & $1.8 \times {10}^{-6}$ \\
n3245 &S0 &   0.44 &0.44 & -19.3 & 4.83 & 0.96 &  0 &  37 &  0.14 & -- &   0.0062 \\
n5866 &S0 &   0.33 &0.30&  -19.3 & 4.54 &1.09 & 0 &  27 & 0.7 & -- &     0.0398 \\
n4262 &SB0&   0.49 &0.55&  -18.4 & 2.13 &0.41 & 2 &  16 & 0.013 & 0.15&  0.0013\\
n3941 &SB0/a& 0.43 &0.43 & -19.2 & 4.39 &0.88 & 1 &  33 & 0.15 & 0.23 &  0.0071\\
n1357 &Sa  &  0.20& 0.23 & -20.2 & 9.78 &2.81 &  4 &  57 & 3.3 &  1.7 &   0.0952\\
n2775 &Sa &   0.33& 0.30&  -19.9 & 7.90 &1.89 &  1.6& 50 & 1.2 & 0.63 &   0.0398\\
n3623 &Sa &   0.35 &0.33 & -20.6 & 15.36 &3.47  & 1.5 &110 & 1.8 & 1.1 &   0.0270\\
n3368 &Sab&   0.27 &0.27 & -20.5 & 13.42 &3.45 &  4   &87  &3.1&  2.0 &    0.0582\\
n3147 &Sb &     --&  -&   -22.0 &  -- &    3.00 &  --& -- & -- &         --\\
n3627 &Sb &   0.14 &0.20 & -19.9 & 7.18 &2.26 &  5 &  38 & 3.1 & 1.6   &  0.1360\\
n1832 &SBb & -0.08 &0.10 & -20.6 & 12.09 &5.17 &  17 & 63 & 12.7&  10.3  & 0.4047\\
n5248 &Sbc &  0.02 &0.11 & -20.3 & 9.29 &3.86 &  9  & 45  &8.5 & 4.4  &   0.3657\\
n6217 &SBbc& -0.22& $<0.001$ &-20.2 &7.51& 4.07&  12 & --&  -- & 6.0 &     0.9992\\
n2903 &Sc &   0.00& 0.11 & -20.1 & 7.73 &3.21 &  8 &  28 & 5.3 & 3.3 &    0.3657\\
n4631 &Sc &     --  &--&   -20.5 &  -- &          40 & --&  -- & -- &        --\\
n6181 &Sc &  -0.13 &0.03 & -20.7 & 12.26 &6.25 &  15 & 75 & 13.9&  11.1 &  0.7791\\
n6643 &Sc &  -0.13 &0.03 & -20.7 & 12.26 &6.25 &  11 & 75 & 13.9 & 8.4 &   0.7791\\
n4775 &Sc  &    -- & --  & -19.9 &  -- &          35 & -- & -- & --  &       --\\
\noalign{\smallskip}
\hline
\end{tabular}

\end{flushleft}

\end{table}

\clearpage

\BTA

\caption[]{}

\begin{flushleft}

\TAB{lllll}
\hline
\noalign{\smallskip}
Age& Sa & Sb &Sc  & Sd  \\
\noalign{\smallskip}
\hline
\noalign{\smallskip}
  0.00E+00 & 0.60 &      0.45         &  0.25       & 0.10\\
  2.50E+08 & 1.57 &      0.98	  &  0.47       & 0.18\\
  5.00E+08 & 2.44 &      1.47	  &  0.69       & 0.27\\
  7.50E+08 & 3.16 &      1.91	  &  0.90       & 0.36\\
  1.00E+09 & 3.72 &      2.28	  &  1.08       & 0.44\\
  1.25E+09 & 4.10 &      2.59	  &  1.25       & 0.52\\
  1.50E+09 & 4.27 &      2.81	  &  1.39       & 0.59\\
  1.75E+09 & 4.24 &      2.95	  &  1.51       & 0.66\\
  2.00E+09 & 3.97 &      3.00	  &  1.60       & 0.72\\
  2.25E+09 & 3.61 &      2.95	  &  1.66       & 0.78\\
  2.50E+09 & 3.28 &      2.80	  &  1.69       & 0.83\\
  2.75E+09 & 2.99 &      2.61	  &  1.68       & 0.87\\
  3.00E+09 & 2.73 &      2.44	  &  1.64       & 0.90\\
  3.25E+09 & 2.49 &      2.28	  &  1.58       & 0.92\\
  3.50E+09 & 2.27 &      2.13	  &  1.52       & 0.93\\
  3.75E+09 & 2.07 &      1.99	  &  1.46       & 0.94\\
  4.00E+09 & 1.89 &      1.86	  &  1.41       & 0.93\\
  4.25E+09 & 1.73 &      1.73	  &  1.35       & 0.91\\
  4.50E+09 & 1.58 &      1.62	  &  1.30       & 0.90\\
  4.75E+09 & 1.45 &      1.52	  &  1.26       & 0.89\\
  5.00E+09 & 1.32 &      1.42	  &  1.21       & 0.87\\
  5.24E+09 & 1.21 &      1.32	  &  1.17       & 0.86\\
  5.50E+09 & 1.10 &      1.24	  &  1.12       & 0.85\\
  5.75E+09 & 1.01 &      1.16	  &  1.08       & 0.84\\
  6.00E+09 & 0.92 &     1.08	  &  1.04       & 0.82\\
  6.24E+09 & 0.85 &     1.01	  &  1.00       & 0.81\\
  6.50E+09 & 0.77 &     0.95	  &  0.97       & 0.80\\
  6.75E+09 & 0.71 &     0.89         &  0.93       & 0.79\\
  7.00E+09 & 0.65 &     0.83	  &  0.90       & 0.78\\
  7.24E+09 & 0.59 &     0.78	  &  0.87       & 0.76\\
  7.50E+09 & 0.54 &     0.73	  &  0.84       & 0.75\\
  7.75E+09 & 0.50 &     0.68	  &  0.81       & 0.74\\
  8.00E+09 & 0.45 &     0.64	  &  0.78       & 0.73\\
  8.24E+09 & 0.42 &     0.60	  &  0.75       & 0.72\\
  8.50E+09 & 0.38 &     0.56	  &  0.72       & 0.71\\
  8.75E+09 & 0.35 &     0.52	  &  0.69       & 0.70\\
  9.00E+09 & 0.32 &     0.49	  &  0.67       & 0.69\\
  9.24E+09 & 0.29 &     0.46	  &  0.65       & 0.68\\
  9.50E+09 & 0.27 &     0.43	  &  0.62       & 0.67\\
  9.74E+09 & 0.24 &     0.40	  &  0.60       & 0.66\\
  1.00E+10 & 0.22 &     0.37	  &  0.58       & 0.65\\
  1.02E+10 & 0.20 &     0.35	  &  0.56       & 0.64\\
  1.05E+10 & 0.19 &     0.33	  &  0.54       & 0.63\\
  1.07E+10 & 0.17 &     0.31	  &  0.52	& 0.62\\
  1.10E+10 & 0.16 &     0.29	  &  0.50	& 0.61\\
  1.12E+10 & 0.14 &     0.27	  &  0.48	& 0.60\\
  1.15E+10 & 0.13 &     0.25	  &  0.46	& 0.60\\
  1.17E+10 & 0.12 &     0.24	  &  0.45	& 0.59\\
  1.20E+10 & 0.11 &     0.22	  &  0.43	& 0.58\\
  1.22E+10 & 0.10 &     0.21	  &  0.42	& 0.57\\
  1.25E+10 & 9.57E-02 & 0.19         &  0.40       & 0.56\\
  1.27E+10 & 8.78E-02 & 0.18	  &  0.39	& 0.55\\
  1.30E+10 & 8.05E-02 & 0.17	  &  0.37	& 0.55\\
  1.32E+10 & 7.38E-02 & 0.16	  &  0.36	& 0.54\\
  1.35E+10 & 6.77E-02 & 0.15	  &  0.35	& 0.53\\
  1.37E+10 & 6.21E-02 & 0.14         &  0.33	& 0.52\\
  1.40E+10 & 5.70E-02 & 0.13         &  0.32	& 0.52\\
  1.42E+10 & 5.22E-02 & 0.12         &  0.31	& 0.51\\
  1.45E+10 & 4.79E-02 & 0.11         &  0.30	& 0.50\\
  1.47E+10 & 4.39E-02 & 0.11         &  0.29	& 0.49\\
  1.50E+10 & 4.03E-02 & 0.10         &  0.28	& 0.49\\
  1.52E+10 & 3.70E-02 & 9.71E-02      &  0.27	& 0.48\\
  1.55E+10 & 3.39E-02 & 9.10E-02      &  0.26	& 0.47\\
  1.57E+10 & 3.11E-02 & 8.53E-02      &  0.25	& 0.47\\
  1.60E+10 & 2.85E-02 & 7.99E-02      &  0.24	& 0.46\\
\noalign{\smallskip}	   		  		
\hline			   
\ETAB			   
			   
\end{flushleft}		   
			   
\ETA			   

\clearpage

\BTA

\caption[]{}

\begin{flushleft}

\TAB{lcccc}
\hline
\noalign{\smallskip}
Type & $<$Z$>$ & Z & SFR & $M_g$/$M_{tot}$ \\
\noalign{\smallskip}
\hline
\noalign{\smallskip}
Sa          & $1.13 \times 10^{-2}$  & $ 5.41 \times 10^{-2}$   & 0.03 & $5.92 \times 10^{-3}$ \\
Sb          & $2.39 \times 10^{-2}$  & $ 4.01 \times 10^{-2}$    & 0.80 & 0.02 \\
Sc          & $1.81 \times 10^{-2}$  & $ 2.20 \times 10^{-2}$   & 2.45 & 0.11 \\
Sd          & $6.13 \times 10^{-3}$  & $ 8.61 \times 10^{-3}$   & 4.57 & 0.43 \\
\noalign{\smallskip}
\hline
\ETAB

\end{flushleft}

\ETA

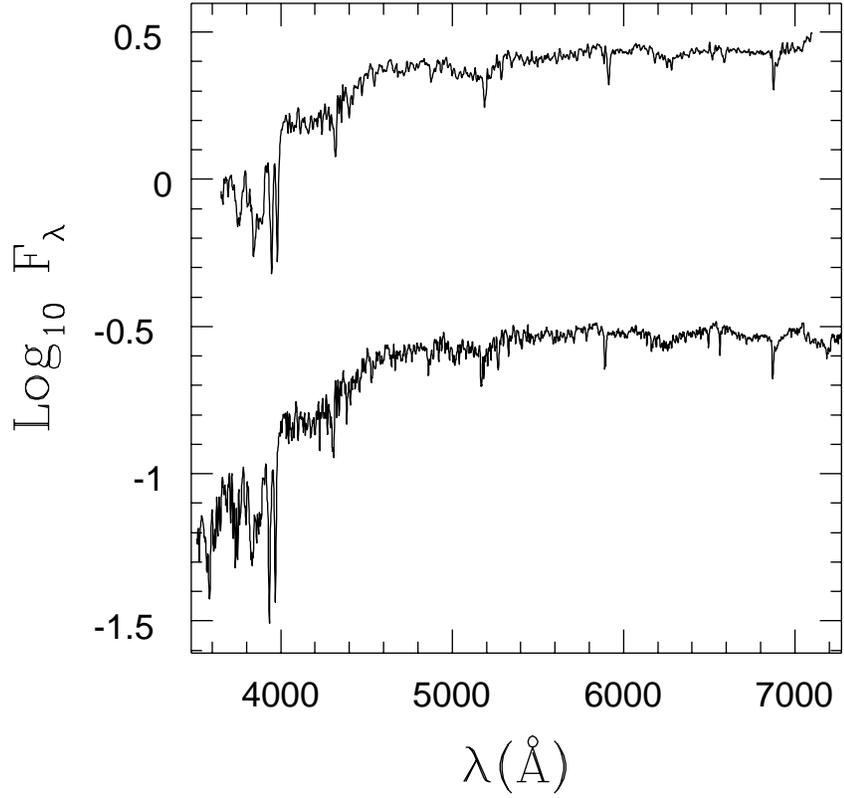
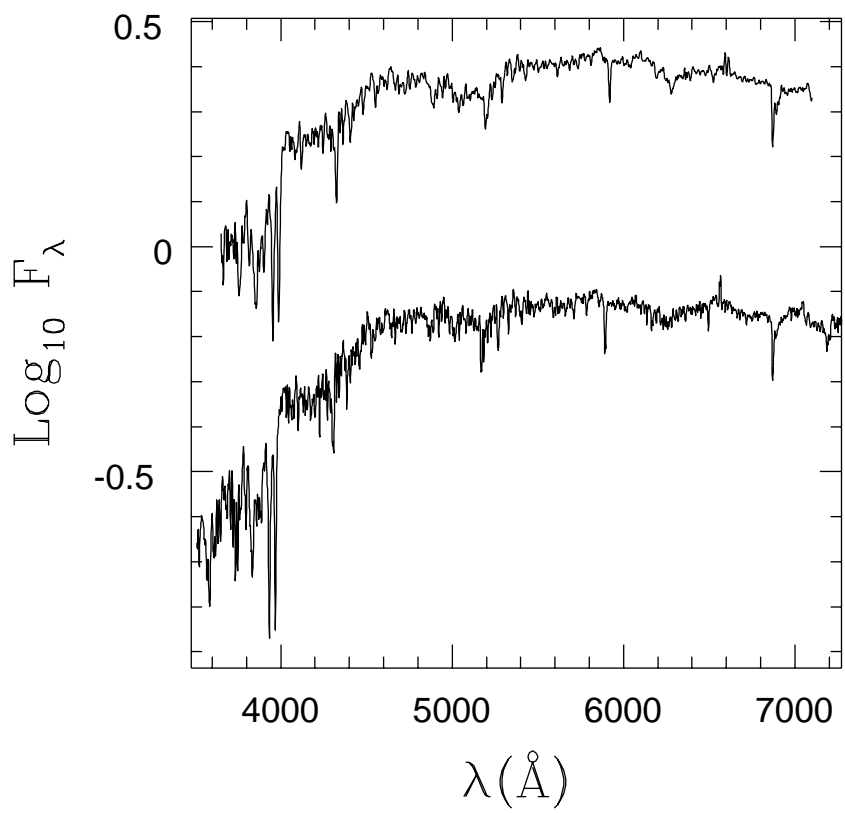
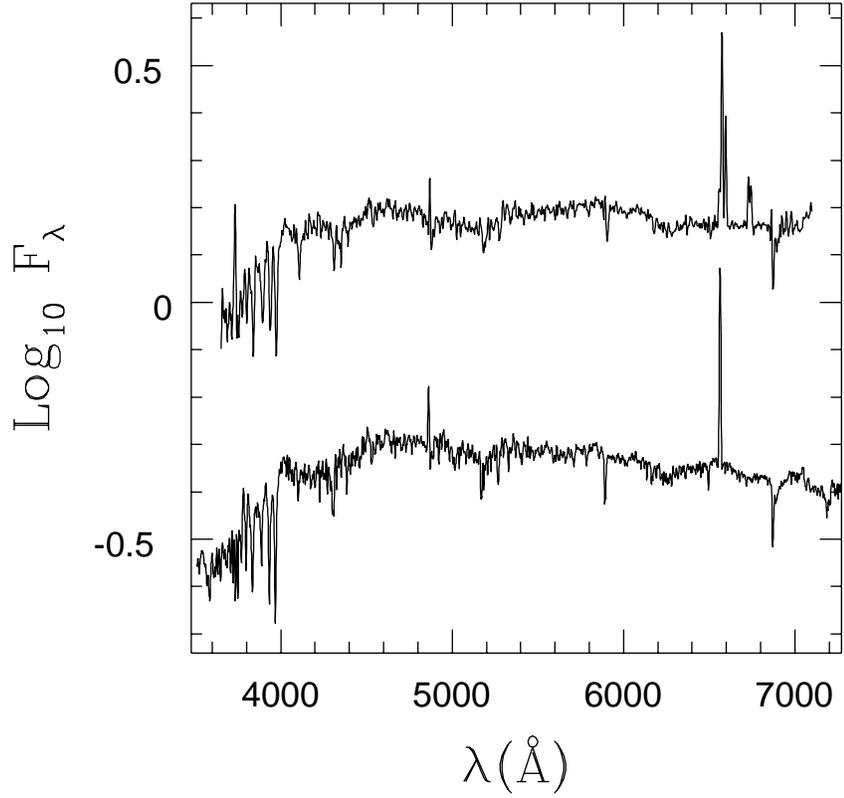
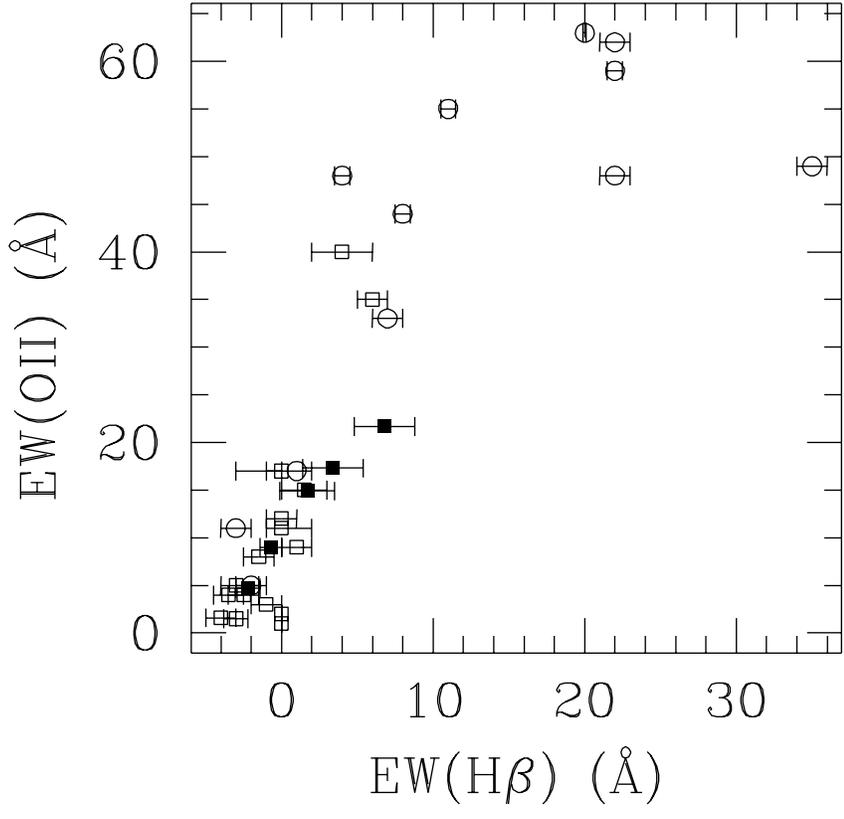
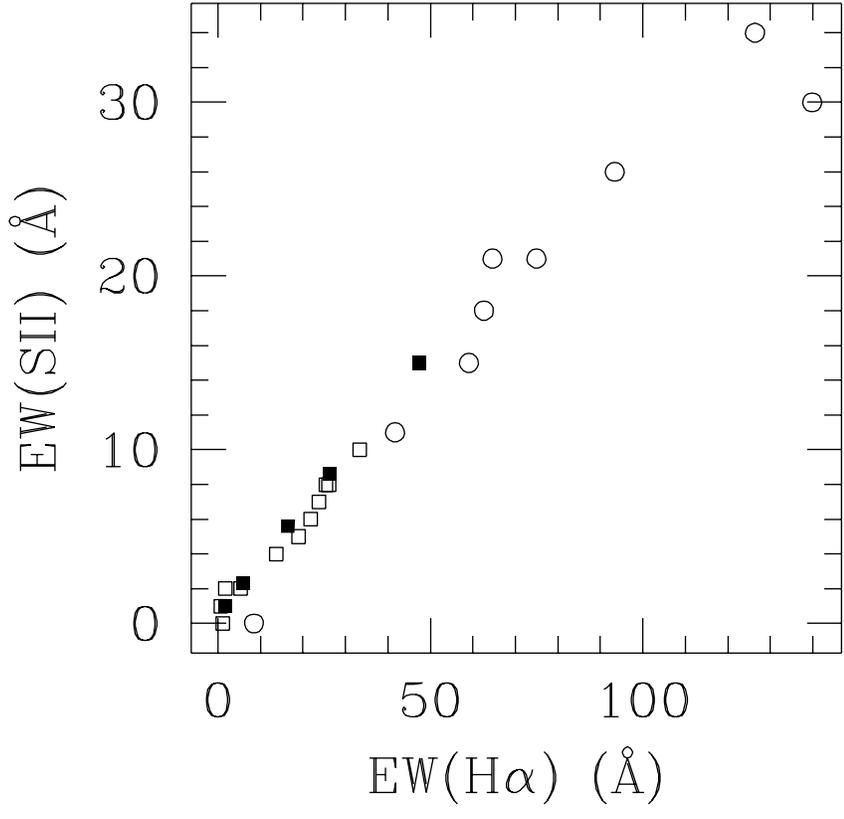
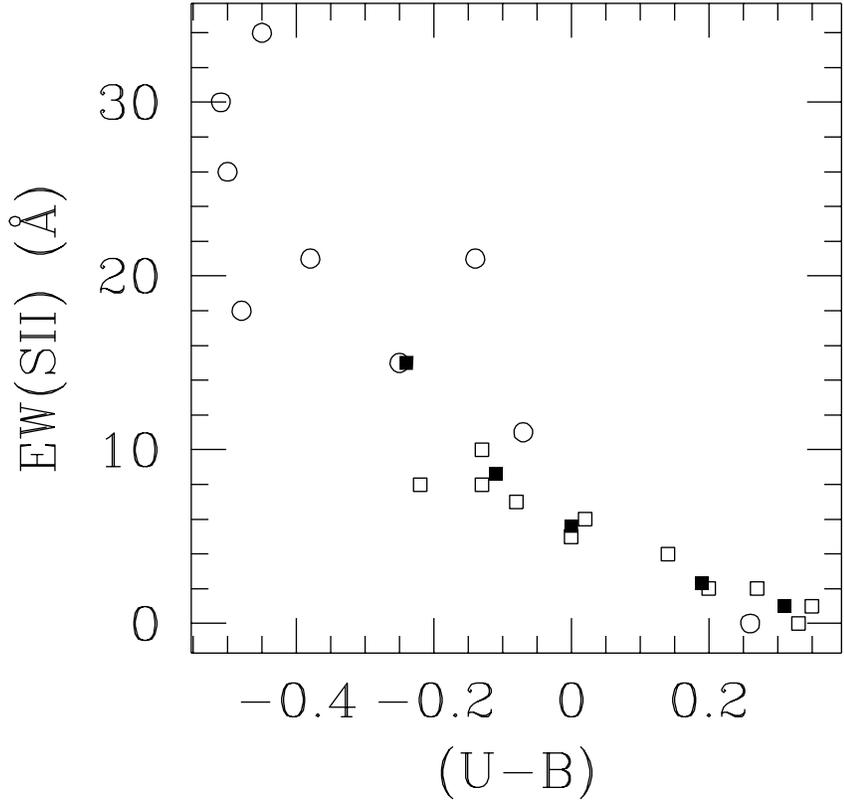
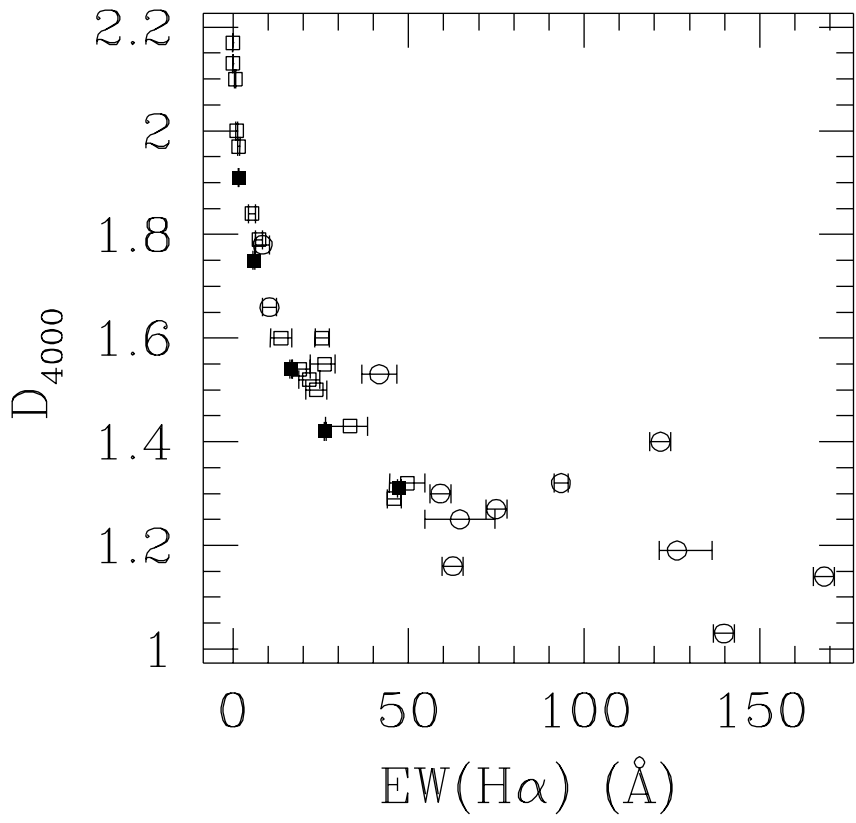
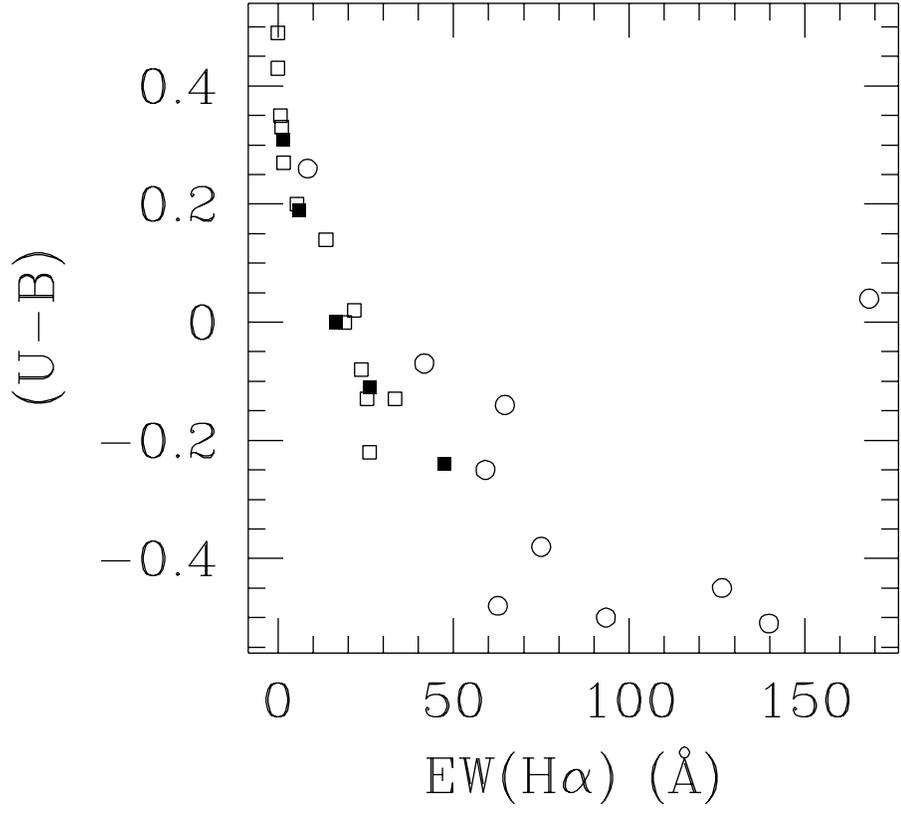
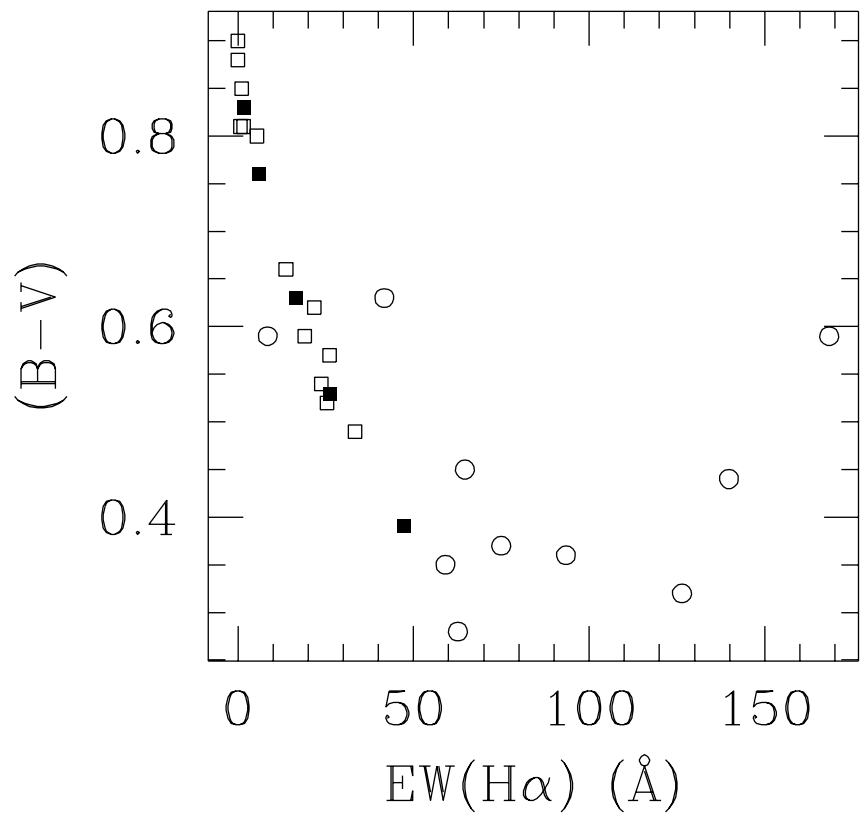
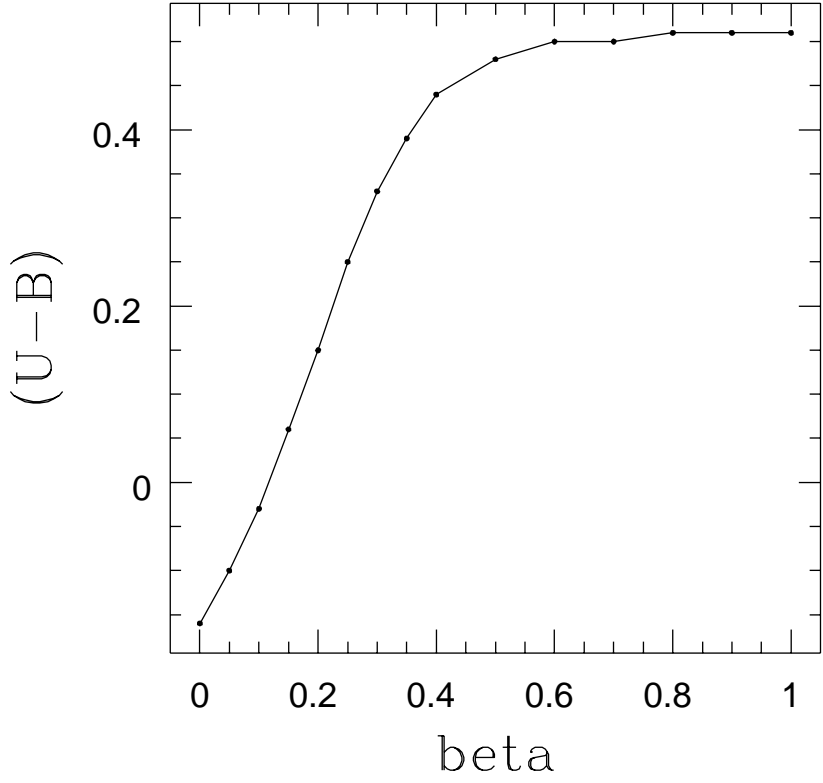
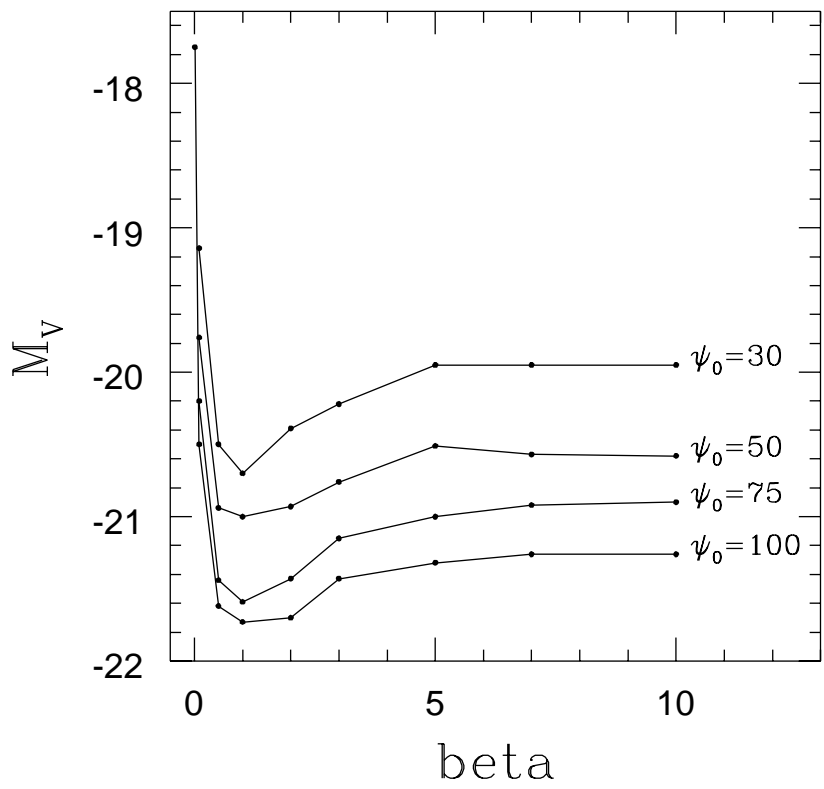
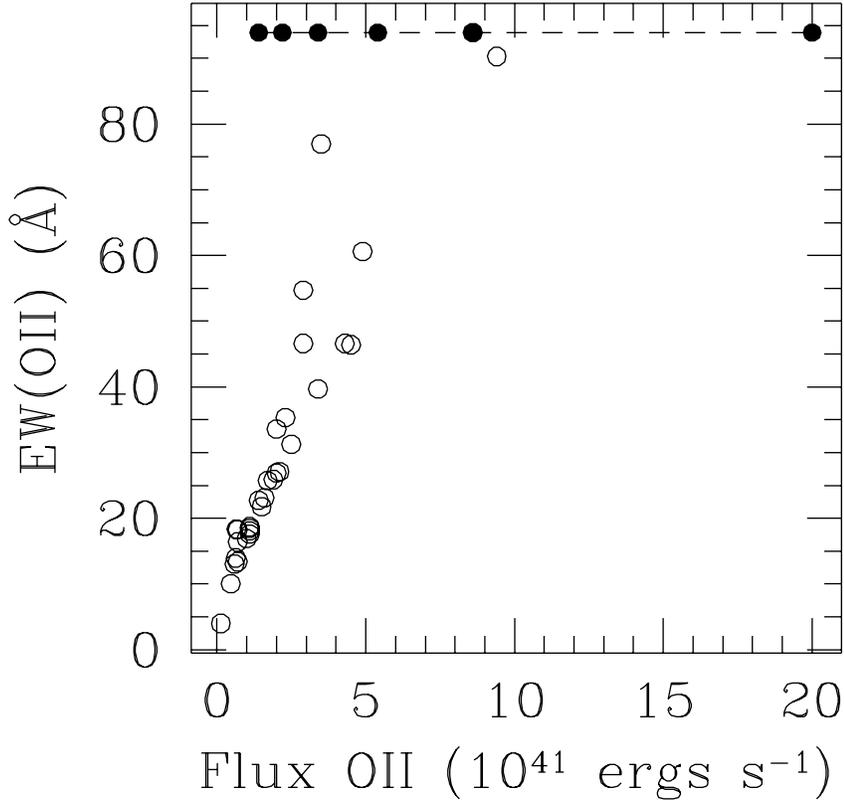
\end{document}